\documentclass[twocolumn,showpacs,preprintnumbers,amsmath,amssymb, prx, reprint, longbibliography]{revtex4-1}
\usepackage{amsmath}
\usepackage{stmaryrd}
\usepackage{txfonts}
\usepackage{amssymb}
\usepackage{mathrsfs}
\usepackage{graphicx}
\usepackage{dcolumn}
\usepackage{bm}
\usepackage{epsfig}
\usepackage{color}
\usepackage{pbox}
\usepackage[colorlinks=true, linkcolor=blue, citecolor=blue, urlcolor=blue]{hyperref} 
\begin{document}
\preprint{\href{https://doi.org/10.1103/PhysRevB.96.014407}{S.-Z. Lin, Phys. Rev. B {\bf 96}, 014407 (2017).}}

\title{Dynamics and inertia of a skyrmion in chiral magnets and interfaces: A linear response approach based on magnon excitations}
\author{Shi-Zeng Lin}
\affiliation{Theoretical Division, T-4 and CNLS, Los Alamos National Laboratory, Los Alamos, New Mexico 87545, USA}

\begin{abstract}
Taking all the magnon modes into account, we derive the skyrmion dynamics in response to a weak external drive. A skyrmion has rotational symmetry and the magnon modes can be characterized by an angular momentum. For a weak distortion of a skyrmion, only the magnon modes with an angular momentum $|m|=1$ govern the dynamics of skyrmion topological center. The skyrmion inertia is determined by the magnon modes in the continuum spectrum. For a skyrmion driven by a magnetic field gradient or by a spin transfer torque generated by a current, the dynamical response is practically instantaneous. This justifies the rigid skyrmion approximation used in Thiele's collective coordinate approach. For a skyrmion driven by a spin Hall torque, the torque couples to the skyrmion motion through the magnons in the continuum and damping, therefore the skyrmion dynamics shows sizable inertia in this case. The trajectory of a skyrmion is an ellipse for an ac drive of spin Hall torque.
\end{abstract}
\date{\today}
\maketitle

\section{Introduction}

A magnetic skyrmion in magnets is swirling spin texture that behaves as a particle with a long lifetime \cite{Skyrme61,Skyrme1962556,Bogdanov89} due to the topological protection. They were observed experimentally recently in chiral magnets where the inversion symmetry is broken. \cite{Muhlbauer2009,Yu2010a}  For their unique topological properties and long lifetime, skyrmions have attracted considerable interests as possible information carriers. Skyrmions can be driven by various external fields, such as electric current \cite{Jonietz2010,Yu2012,Schulz2012}, magnetic/electric field gradient \cite{White2012,PhysRevLett.113.107203}, thermal gradient \cite{Kong2013,Lin2014PRL,Mochizuki2014} and magnon current \cite{PhysRevB.90.094423} etc. The ability to manipulate skyrmions by an electric current is especially attractive because this implies immediately that skyrmions can be used in spintronic devices. \cite{Fert2013,nagaosa_topological_2013} Moreover the threshold current to make skyrmions mobile is weak, thanks to the smooth spin texture associated with the skyrmion. 

For applications, it is crucial to understand the dynamics of skyrmion in response to an external drive. The equation of motion of a rigid skyrmion in two dimensions (in the $x$-$y$ plane) has been obtained by Thiele long time ago. \cite{Thiele72} It has the following form
\begin{align}\label{eq0}
\mathbf{G}_T\times \mathbf{v}+D_T \mathbf{v}=\mathbf{F}_T,
\end{align}
where $\mathbf{v}$ is the skyrmion velocity. The first term describes gyromotion with the gyrovector $\mathbf{G}_T$ perpendicular to the $x$-$y$ plane, and the second term is the damping. Here $D_T\ll G_T$ and the skyrmion moves almost perpendicular to the external force $\mathbf{F}_T$. In the Thiele's collective coordinate approach, the skyrmion texture $\mathbf{S}(\mathbf{r})$ is assumed to be rigid and it moves as a whole, $\mathbf{S}[\mathbf{r}-\mathbf{R}(t)]$, with $\mathbf{R}(t)$ representing the translational motion of a skyrmion. It has been demonstrated in numerical simulations that the Thiele's equation of motion correctly captures the skyrmion dynamics driven by a spin transfer torque induced by a dc current. \cite{Iwasaki2013,szlin13skyrmion1,szlin13skyrmion2} Nevertheless, it is unknown to date how to justify the rigid skyrmion approximation.
  
A skyrmion is a collective excitation of spin texture and thus has internal degrees of freedom. \cite{Lin_internal_2014,PhysRevB.90.094423} In the presence of an external drive, the skyrmion can be deformed when the magnon modes are excited by the external drive. In the continuum approximation where the system preserves the translational symmetry, there is a Goldstone mode in the characteristic deformations of the skyrmion corresponding to the translational motion of a skyrmion. \cite{Everschor11,Everschor12,PhysRevB.58.R8889} The Thiele's equation of motion includes only the Goldstone mode. In principle, the deformations associated with other magnon modes can also be involved in the motion of a skyrmion. To go beyond the Thiele's approach, one needs to take all the magnon excitations into account. One can introduce corrections to the Thiele's equation of motion, such as mass and gyrodamping, and then fit the generalized Thiele's equation of motion to the skyrmion trajectory obtained from direct simulation of the spin dynamics based on the Landau-Lifshitz-Gilbert equation. In this way the parameters in the Thiele equation of motion can be extracted. \cite{PhysRevB.50.12711,PhysRevB.90.174434,buttner_dynamics_2015} Similar to the original Thiele's equation of motion, the generalized Thiele's equation in prior is not justified. A theory to describe the skyrmion dynamics by treating all the magnon modes on an equal footing thus is required. 
  
In this work, we present a linear theory for the skyrmion dynamics by taking all the magnon modes into account. Specifically, we consider a small oscillation of a skyrmion subjected to a weak oscillating magnetic field gradient, spin transfer torque and spin Hall torque. To do that, we first define the skyrmion center as its topological charge center. We then express the dynamics of the skyrmion center in terms of the magnon modes. We find that only the magnon modes with an angular momentum $|m|=1$ are responsible for the dynamics of skyrmion center. The retardation or inertia of skyrmions is due to the magnon modes in the continuum. For a skyrmion driven by a field gradient or spin transfer torque, the inertia is negligible, which justifies the Thiele's rigid skyrmion approximation. For a spin Hall torque, the motion of skyrmions is originated from the coupling between the magnon continuum and current. The inertia is significant in this case. The inertia can be quantified by measuring the phase shift between the skyrmion velocity and the driving field.

\section{Model and Bloch skyrmion solution}\label{Sec2}

We consider the following Hamiltonian density for spins $\mathbf{S}(\mathbf{r})$ in two dimensional space \cite{Bogdanov89}
\begin{equation}\label{eq1}
{\cal H} = \frac{{{J_{\mathrm{ex}}}}}{2}\mathop \sum \limits_{\mu  = x,y} {\left( {{\partial _\mu }{\bf{S}}} \right)^2} + D{\bf{S}}\cdot\nabla  \times {\bf{S}} - {\bf{B}}\cdot{\bf{S}},
\end{equation}
which successfully captures many experimental observations in chiral magnets. Here $\mathbf{S}$ is a unit vector representing the spin direction, $J_{\mathrm{ex}}$ is the exchange interaction, $D$ is the Dzyaloshinskii-Moriya (DM) interaction \cite{Dzyaloshinsky1958,Moriya60,Moriya60b} and $\mathbf{B}=B\hat{z}$ with a unit vector $\hat{z}$ is the external magnetic field perpendicular to the plane. We have neglected the weak dipolar interaction. The skyrmion size is much bigger than the spin lattice constant, and this justifies the continuum approximation in Eq. \eqref{eq1}. We renormalize the length in unit of $J_{\mathrm{ex}}/D$ and energy density and $B$ in unit of $D^2/J_{\mathrm{ex}}$. Then Eq. \eqref{eq1} takes a dimensionless form. We will use dimensionless quantities in the following derivations. The Hamiltonian Eq. \eqref{eq1} supports the Bloch skyrmion solution. The results for the N\'{e}el skyrmions will be discussed in Sec. VI.

\begin{figure}[b]
\psfig{figure=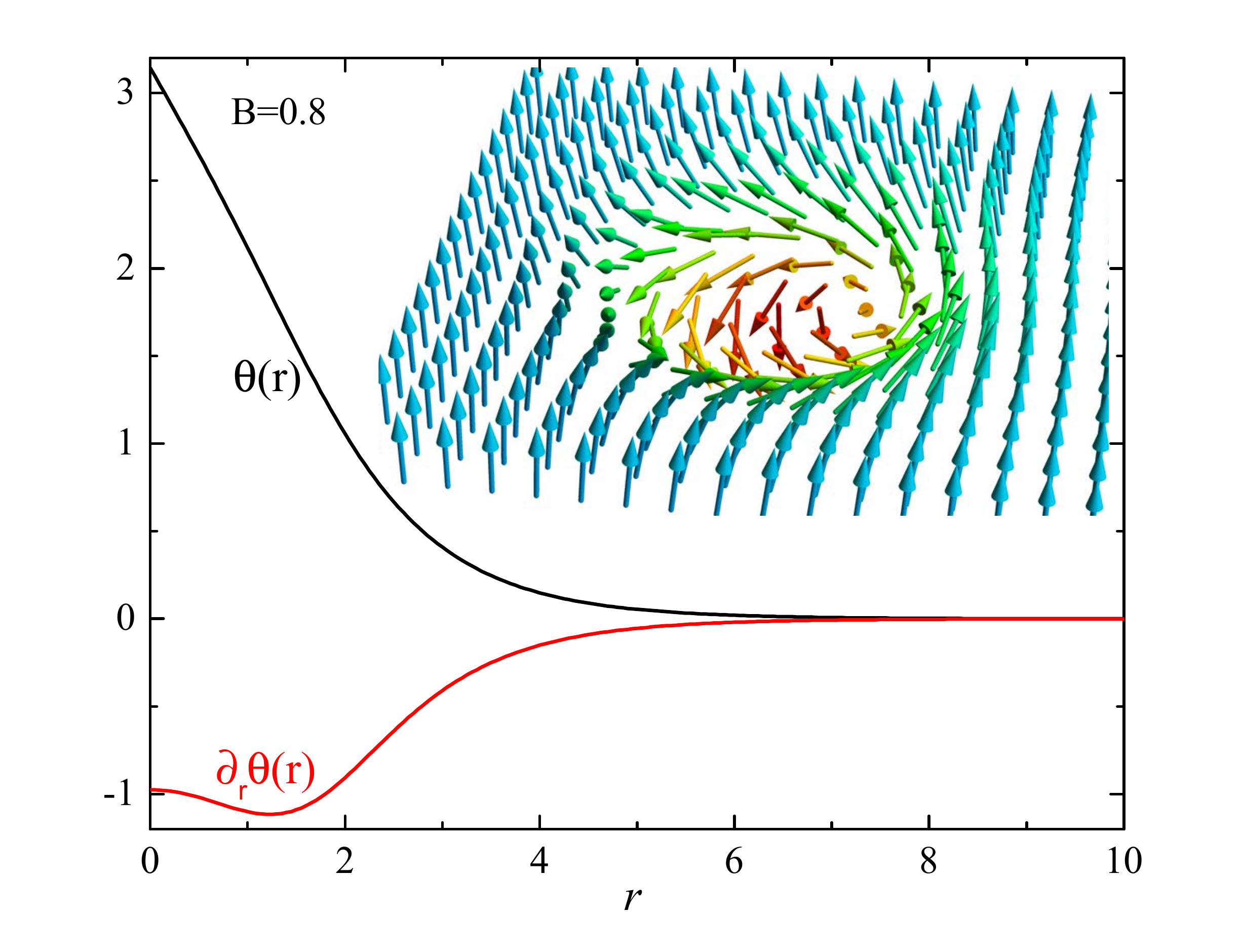,width=\columnwidth}
\caption{(color online) Profiles of $\theta(r)$ and $\partial_r\theta(r)$ for a skyrmion in the ferromagnetic background. Inset is a schematic view of a skyrmion. Here $B=0.8$.
} \label{f1}
\end{figure}

We focus on a single skyrmion in the ferromagnetic background, which is a metastable state for $B>0.55$. \cite{Bogdanov94,Lin_internal_2014,PhysRevB.90.094423} Because of the rotational symmetry, it is more convenient to work in the polar coordinate $\mathbf{r}=(r,\  \phi)$. The skyrmion solution is $\mathbf{S}_0=(\cos\varphi\sin\theta,\ \sin\varphi\sin\theta, \ \cos\theta)$ with $\varphi=\phi+\pi/2$ and $\theta(r)$ being a function of $r$ only. The stationary skyrmion solution is obtained by minimizing $\mathcal{H}$ with respect to $\theta$, and we have equation for $\theta(r)$
\begin{equation}\label{eq2}
\cos \left( {2\theta } \right) + \frac{1}{{2r}}\sin(2\theta)  + B r\sin\theta  - \left( {{\partial _r}\theta  + 1} \right) - r\partial _r^2\theta  = 0.
\end{equation}
We solve Eq. \eqref{eq2} using the relaxation method to find $\theta(r)$. The skyrmion structure and the results for $\theta(r)$ and $\partial_r\theta(r)$ at $B=0.8$ are displayed in Fig. \ref{f1}.

\section{Eigenmodes analysis}
We calculate the eigenmodes of a skyrmion in the ferromagnetic background, following the approach in Ref. \onlinecite{PhysRevB.90.094423}. We introduce a local coordinate system with the local $z$ axis along the spin direction $\mathbf{S}$. The spin representation in the lab coordinate and the local coordinate are related by the following subsequent rotation operations in the lab frame: rotation along the $z$ axis by $\phi_0$, rotation along the $y$ axis by $\theta$ and rotation along the $z$ axis by $\varphi$.  We choose $\phi_0=\pi/2$. Then the spin in the local coordinate $\mathbf{L}=(L_X,\ L_Y,\ L_Z)$ is related to that in the lab frame $\mathbf{S}$ according to $\mathbf{S}=\hat{O}\mathbf{L}$, where \cite{Lin_internal_2014}
\begin{equation}\label{eq4}
\hat{O}= \left(
\begin{array}{ccc}
 -\text{sin$\varphi $} & -\text{cos$\varphi $} \text{cos$\theta $} & \text{cos$\varphi $} \text{sin$\theta $} \\
 \text{cos$\varphi $} & -\text{sin$\varphi $} \text{cos$\theta $} & \text{sin$\varphi $} \text{sin$\theta $} \\
 0 & \text{sin$\theta $} & \text{cos$\theta $} \\
\end{array}
\right).
\end{equation}
In the stationary state $\bar{L}_{X}=\bar{L}_Y=0$ and $\bar{L}_Z=1$. We consider small deviations $\mathbf{L}$ from the stationary state and introduce the magnon fields
\begin{equation}\label{eq5}
\psi=\frac{L_X+i L_Y}{\sqrt{2}}, \ \ \ \psi^*=\frac{L_X-i L_Y}{\sqrt{2}},
\end{equation}
and $L_Z=1-\psi\psi^*$ with $|\psi|\ll 1$. Expanding the Hamiltonian to the second order in $\psi$, we obtain
\begin{equation}\label{eq6}
\mathcal{H}_\psi=\frac{1}{2}\hat{\psi}^\dagger\mathcal{H}_\psi\hat{\psi},
\end{equation}
\begin{equation}\label{eq7}
\mathcal{H}_\psi=(-\nabla^2+V_0)I_0-2\sigma_z\left(\frac{\cos\theta}{r^2}-\frac{\sin\theta}{r}\right)i\partial_\phi+V_1\sigma_x,
\end{equation}
with $\sigma_i$ ($i=x,\ y,\ z$) being the Pauli matrices and $I_0$ the unit matrix. Here
\begin{equation}\label{eq8}
\hat{\psi}^\dagger=(\psi^*,\ \  \psi),
\end{equation}
\begin{equation}\label{eq9}
{V_0} = \frac{{1 + 3\cos \left( {2\theta } \right)}}{{4{r^2}}} - \frac{{3\sin \left( {2\theta } \right)}}{{2r}} + {B_z}\cos\theta  - {\partial _r}\theta  - \frac{1}{2}{\left( {{\partial _r}\theta } \right)^2},
\end{equation}
\begin{equation}\label{eq10}
{V_1} = \frac{{{{\sin }^2}\theta }}{{2{r^2}}} + \frac{{\sin \left( {2\theta } \right)}}{{2r}} - {\partial _r}\theta  - \frac{1}{2}{\left( {{\partial _r}\theta } \right)^2}.
\end{equation}
Equations \eqref{eq6}-\eqref{eq10} were first derived in Ref. \onlinecite{PhysRevB.90.094423}. Note that with the definition of magnon wave function in Eq. \eqref{eq5}, there is a minus sign in front of $\sigma_z$ in Eq. \eqref{eq7} which is different from the results in Ref. \onlinecite{PhysRevB.90.094423}. The eigenmodes are determined by the equation
\begin{equation}\label{eq11}
-i\sigma_z\partial_t\hat{\psi}=\mathcal{H}_\psi\hat{\psi}.
\end{equation}
This equation for magnons has the form of the Schr\"{o}dinger equation in the presence of a centrosymmetric potential. We can introduce an angular momentum $m$ with $\psi=\psi_m(r,t)\exp(i m\phi)$ to classify the eigenmodes. The two components in $\hat{\psi}$ are related by complex conjugation because the magnetic moment $\mathbf{S}$ is real. This indicates that the matrix equation in Eq. \eqref{eq11} is reductant. Indeed $\mathcal{H}_\psi$ has particle-hole symmetry, $\mathcal{H}_\psi=\sigma_x K\mathcal{H}_\psi K \sigma_x$ with $K$ being the complex conjugate operator. This means that if $\exp[i(\omega t+m\phi)]\hat{\eta}_m$, with $\hat{\eta}_m^\dagger\equiv (\eta_1^*, \ \ \eta_2^*)$, solves Eq. \eqref{eq11}, then $\exp[-i(\omega t+m\phi)]\sigma_x K\hat{\eta}_m$ also solves Eq. \eqref{eq11}. Then $\hat{\psi}$ can be obtained by linear superposition of the two symmetry-related solutions
\begin{equation}\label{eq12}
\hat{\psi}_m=b \exp[i(\omega t+m\phi)]\hat{\eta}_m+b^*\exp[-i(\omega t+m\phi)]\sigma_x K\hat{\eta}_m.
\end{equation}
For $\hat{\psi}_m$ in Eq. \eqref{eq12}, its two components are complex conjugate to each other. Here $\hat{\eta}$ is determined by the eigenvalue problem
 \begin{equation}\label{eq13}
\omega_m\sigma_z\hat{\eta}_m=\mathcal{H}_\psi\hat{\eta}_m.
\end{equation}
To solve Eq. \eqref{eq13}, we use the Bessel functions $J_m(k r)$ as basis to represent the matrix $\mathcal{H}_\psi$ as detailed in the Appendix A. From $\hat{\eta}_m$, we know $\mathbf{L}_m$ from Eq. \eqref{eq5}. Then we obtain the eigenmodes $\tilde{\mathbf{S}}_m$ in the lab frame through rotation, $\tilde{\mathbf{S}}_m=\hat{O}\mathbf{L}_m$.

In the high energy limit $\omega\gg \omega_g$, with $\omega_g=B$ being the magnon gap, we can use magnon momentum $k$ to label the magnon mode. The eigenfrequency in this limit is $\omega_m(k)=k^2+\omega_g$ and the eigenmodes are $\hat{\eta}_m^\dagger=(1,\ 0)J_m(k r)$. When $\omega$ is comparable to $\omega_g$, the momentum $k$ is not a good quantum number because the presence of the skyrmion breaks the translational invariance for magnons.

In the continuum limit adopted in Eq. \eqref{eq1}, valid when the skyrmion size is much bigger than the spin lattice constant, the mode corresponding to the translational motion of a skyrmion is a Goldstone mode. It has $|m|=1$ and the magnon mode in the lab frame is
\begin{align}\label{eq14}
{\tilde{\mathbf{S}}_{|m| = 1,j=1}} = \frac{{{{i}m}{\partial _x}{\mathbf{S}_0} - {\partial _y}{\mathbf{S}_0}}}{{\sqrt 2 \sqrt {\int{{d}}{r^2}{{\left( {{\partial _x}{\mathbf{S}_0}} \right)}^2}} }}.
\end{align}
The imaginary and real part of ${\tilde{\mathbf{S}}_{|m| = 1}}$ correspond to the translation motion along the $x$ and $y$ direction respectively.

\section{Linear response to external drive}
We proceed to calculate the response of skyrmion to external drive in terms of the eigenmodes. We consider the Landau-Lifshitz-Gilbert equation of motion for $\mathbf{S}$
 \begin{equation}\label{eq15}
\partial_t\mathbf{S} =-\mathbf{S}\times \mathbf{H}_{\mathrm{eff}}+\alpha  \mathbf{S}\times \partial_t\mathbf{S} +\mathbf{\Gamma},
\end{equation}
where $\alpha$ is the Gilbert damping, $\mathbf{H}_{\mathrm{eff}}\equiv -\delta\mathcal{H}/\delta\mathbf{S}$ is the effective magnetic field and $\mathbf{\Gamma}$ is the torque due to external fields. Here the time is in unit of $J_{\mathrm{ex}}/(\gamma D^2)$ with $\gamma$ the gyromagnetic ratio. We study the linear response of a skyrmion to a weak torque, $|\mathbf{\Gamma|}\ll 1$. We will work in the lab frame here. We consider a small oscillation of the skyrmion center around the equilibrium position. The linear response $\mathbf{S}=\mathbf{S}_0+\tilde{\mathbf{S}}$ is governed by
\begin{equation}\label{eq16}
{\partial _t}\tilde{\mathbf{S}}(t)  = \hat{\mathcal{H}}_S\tilde{\mathbf{S}} + \alpha {\mathbf{S}_0} \times {\partial _t}\tilde{\mathbf{S}} +\mathbf{ \Gamma}. 
\end{equation}
Here $\hat{\mathcal{H}_S}$ and $\tilde{\mathbf{S}}$ are connected to $\mathcal{H}_\psi$ and $\hat{\psi}$ through rotation, Eq. \eqref{eq4}. Expanding $\tilde{\mathbf{S}}$ in terms of the eigenmodes, we have
\begin{equation}\label{eq17}
\tilde{\mathbf{S}}(t) = \mathrm{Re}\left[\exp(i\omega t)\sum_{m, j}a_{m, j}\tilde{\mathbf{S}}_{m,j}\right].
\end{equation}
The index $j$ labels the modes at a given $m$. Substituting Eq. \eqref{eq17} into Eq. \eqref{eq16} and projecting into the $\tilde{\mathbf{S}}_{m,j}$ mode, we obtain
\begin{align}
a_{m,j}(\omega-\omega_{m,j})+i\omega\sum_{j'}\alpha_{m, j, j'}a_{m,j'}=F_{m,j},
\end{align}
where $F_{m,j}$ represents the coupling between the external torque and the eigenmodes
\begin{equation}
F_{m,j}(\omega) =-i\int dr^2 \tilde{\mathbf{S}}_{m,j}^\dagger\cdot \mathbf{\Gamma},
\end{equation}
and $\alpha_{m, j,j'}$ is the damping coefficient for different modes
\begin{align}\label{eqalphamj}
\alpha_{m,j,j'} =i\alpha \int dr^2 \tilde{\mathbf{S}}_{m,j}^\dagger\cdot\left(\mathbf{S}_0\times \tilde{\mathbf{S}}_{m,j'} \right)\nonumber\\
=i2\pi\alpha\int dr r[L_{Y;m, j}^*(r)L_{X; m,j'}(r)-L_{X;m, j}^*(r)L_{Y;m, j'}(r)].
\end{align}
As shown in the Appendix B, the diagonal elements $|\alpha_{m,j,j}|$ are much larger than the off-diagonal elements $|\alpha_{m,j,j'}|$. For $\alpha\ll 1$, we can neglect the off-diagonal elements of $\alpha_{m,j,j'}$ and take $\alpha_{m,j,j'}=\alpha_{m,j}\delta_{j,j'}$, with $\delta_{j,j'}$ being the Kronecker delta function. For $\omega_{m, j}\gg \omega_g$, we have $\alpha_{m,j}=1$. We then obtain $a_m$ in the frequency domain
\begin{equation}\label{eqamj1}
a_{m,j}(\omega) =\frac{F_{m,j}}{\omega-\omega_{m,j}+i \alpha_{m, j}\omega}.
\end{equation}

To obtain the equation of motion for a skyrmion as a particle, we need to define its center. We use the definition based on its topological charge density
\begin{equation}\label{eq21}
{\bf{R}} (t)= \frac{{\int d{r^2}{\bf{r}}\mathbf{S}\cdot\left( {{\partial _x}\mathbf{S} \times {\partial _y}\mathbf{S}} \right)}}{{\int d{r^2}\mathbf{S}\cdot\left( {{\partial _x}\mathbf{S} \times {\partial _y}\mathbf{S}} \right)}}=\frac{\int d{r^2}{\bf{r}}\mathbf{S}\cdot\left( {{\partial _x}\mathbf{S} \times {\partial _y}\mathbf{S}} \right)}{4\pi N_s},
\end{equation}
where $N_s=-1$ is the skyrmion topological charge for the skyrmion shown in Fig. \ref{f1}, which is invariant with respect to small perturbations.

Taking the time derivative of $\mathbf{R}(t)$ and expanding in terms of the eigenmodes, we obtain
\begin{equation}\label{eq22}
{\mathbf{R}}(\omega)=\sum_{m, j}a_{m,j}(\omega)\mathbf{W}_{m,j}(\omega),
\end{equation}
with
\begin{align}\label{eq23}
\begin{split}
\mathbf{W}_{m,j}=\frac{1}{{4\pi {N_s}}}\int {d}{r^2}{\bf{r}}\left[ {{\tilde S}_{m,k}}\cdot\left( {{\partial _x}{S_0} \times {\partial _y}{S_0}} \right)  \right. \\
\left. + {S_0}\cdot\left( {{\partial _x}{{\tilde S}_{m,k}} \times {\partial _y}{S_0}} \right) + {S_0}\cdot\left( {{\partial _x}{S_0} \times {\partial _y}{{\tilde S}_{m,k}}} \right) \right]\\
=\zeta_{m, j}(\hat{x}+i m \hat{y})\delta_{|m|, 1},
\end{split}
\end{align}
and
\begin{align}
\zeta_{m, j}=\frac{1}{4 N_s}\int dr r \left[ i m L_{X;m, j}\partial_r \theta-\partial_r\left(L_{Y; m, j} \sin\theta \right)\right]\delta_{|m|, 1}.
\end{align}
One important observation is that \emph{only the modes with $|m|=1$ couple to the skyrmion center motion}. The lowest mode with $|m|=1$ is the Goldstone mode corresponding to the translational motion of a rigid skyrmion. Other modes with $|m|=1$ lie in the magnon continuum. For a rigid skyrmion, the response to external drive is instantaneous, i.e. the inertia of a skyrmion is absent. Therefore the inertia of skyrmion is contributed from the excitation of the magnon continuum. Note that the linear analysis in Eq. \eqref{eq16} is valid when $\mathbf{R}(\omega)$ is smaller than the skyrmion size. 

An alternative definition of the skyrmion center that is more relevant to experiments is based on the out-of-plane component of the spin
\begin{align}
{\bf{R}}' = \frac{{\int {d}{r^2}{\bf{r}}\left( {{S_z} - 1} \right)}}{{\int {{d}}{r^2}\left( {{S_z} - 1} \right)}}.
\end{align}
The equation of motion for ${\bf{R}}' $ to the linear order in perturbation has the same expressions as those in Eqs. \eqref{eq22} and \eqref{eq23}, except for $\zeta_{m, j}$,
\begin{align}
\zeta_{m, j}=\frac{\pi}{\int dr^2(S_{0, z}-1)}\int dr r^2 L_{Y; m, j} \sin\theta \delta_{|m|, 1}.
\end{align}
In the present work, we use the skyrmion center defined in Eq. \eqref{eq21}.

The external torque $\mathbf{\Gamma}$ is proportional to some control parameters $\mathbf{P}$, such as electric current density or magnetic field gradient. The response of velocity $\mathbf{v}=i\omega \mathbf{R}(\omega)$ to $\mathbf{P}$ can be expressed in a matrix form
\begin{align}\label{eq25}
\left(
\begin{array}{c}
 v_x \\
 v_y \\
\end{array}
\right)=\left(
\begin{array}{cc}
 \chi _{11} & \chi _{12} \\
 \chi _{21} & \chi _{22} \\
\end{array}
\right) \left(
\begin{array}{c}
 P_x \\
 P_y \\
\end{array}
\right).
\end{align} 
The response is isotropic under the spatial rotation of $\mathbf{P}$ and $\mathbf{v}$, which requires $\chi_{22}=\chi_{11}$ and $\chi_{12}=-\chi_{21}$. Here $\chi_{ij}$ can be complex and the skyrmion trajectory is generally an ellipse. The skyrmion inertia manifests itself in the phase shift between the drive $\mathbf{P}(\omega)$ and velocity $\mathbf{v}(\omega)$. We can define a longitudinal phase shift $\Theta_{\parallel}$ and transverse phase shift $\Theta_{\perp}$,
\begin{align}
\tan\Theta_{\parallel}\equiv \frac{\mathrm{Im}[\chi_{11}]}{\mathrm{Re}[\chi_{11}]}, \ \ \ \tan\Theta_{\perp}\equiv \frac{\mathrm{Im}[\chi_{21}]}{\mathrm{Re}[\chi_{21}]}.
\end{align}
To cast Eq. \eqref{eq25} into the standard Thiele's form, we can introduce a generalized force $\mathbf{f}$ from $\mathbf{P}$, and rewrite Eq. \eqref{eq25} into
\begin{align}\label{eq26}
 \left(
\begin{array}{cc}
 \mathcal{D} +i \omega  m  & \mathcal{G}-i \omega  \mathcal{A} \\
 -  \mathcal{G}+i \omega  \mathcal{A} &  \mathcal{D} +i \omega  m  \\
\end{array}
\right)\left(
\begin{array}{c}
  v_x \\
  v_y \\
\end{array}
\right)=\left(
\begin{array}{c}
   f_x \\
   f_y \\
\end{array}
\right).
\end{align} 
Here $ \mathcal{D}\propto \alpha$ is the damping, $m$ is the mass, $\mathcal{G}$ is the gyrocoupling and $\mathcal{A}\propto \alpha$ is the gyrodamping. All those quantities are frequency dependent. The quantities at the left-hand side of Eq. \eqref{eq26} depend on the definition of the generalized force $\mathbf{f}$ at the right-hand side. In our discussions, we will focus on Eq. \eqref{eq25} since it captures fully the dynamics of a skyrmion.

In the limit of $\omega=0$, only the Goldstone modes with $|m|=1$ and $j=1$ contribute to the summation in Eq. \eqref{eq22}. In this limit, the rigid skyrmion approximation employed in the Thiele's collective coordinate approach becomes exact. From the Goldstone modes in Eq. \eqref{eq14}, we obtain $\alpha_{|m|=1,j=1}=-m\alpha/\kappa$ and $W_{|m=1|, j=1}=-(m i\hat{x}-\hat{y})/\sqrt{8\pi\kappa}$, where the skyrmion form factor $\kappa$ is
\begin{align}
\kappa=\int dr^2(\partial_{\mu} S_0)^2/(4\pi),
\end{align} 
with $\mu=x,\ y$. Here $\kappa$ is of the order of unity, $\kappa\sim 1$. The equation of motion becomes
\begin{align}
\mathbf{v}(\omega)=\sum_{m=\pm1, j=1}\frac{i F_{m, j} (-m i\hat{x}+\hat{y})}{\sqrt{8\pi\kappa}(1-i m\alpha/\kappa)}.
\end{align}

\section{Applications}
In the following, we calculate the response matrix $\chi_{ij}$ for a skyrmion driven by a magnetic field gradient, spin transfer torque and spin Hall torque separately.

\begin{figure}[t]
\psfig{figure=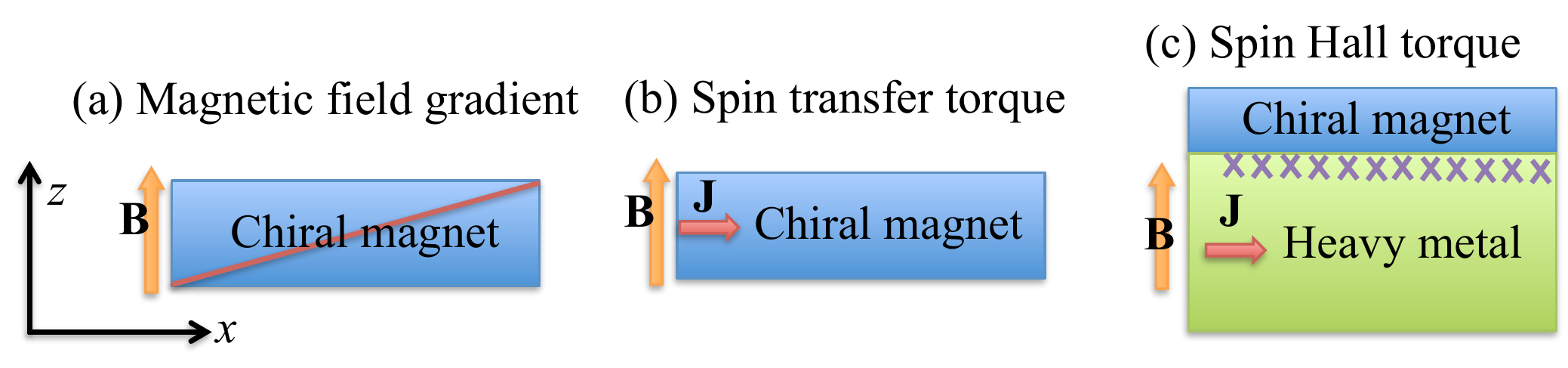,width=\columnwidth}
\caption{(color online) Schematic view of a skyrmion in chiral magnets subjected to different drives: (a) a linear magnetic field gradient, (b) spin transfer torque and (c) spin Hall torque. The crosses in (c) represent the spin accumulation and spin current normal to the interface due to the spin Hall effect.
} \label{f1bb}
\end{figure}

\subsection{Magnetic field gradient}
We first consider the skyrmion motion driven by a magnetic field gradient as shown in Fig. \ref{f1bb} (a). This has been studied by numerical simulations recently \cite{moon_magnetic_2015,moon_skyrmion_2016}. We assume that the field is along the $z$ axis and the gradient is along the $x$ direction. The total magnetic field is $\mathbf{B}_T=B\hat{z}+\Delta_B^{(x)} r\cos\phi \hat{z}$. The magnetic field in the first term stabilizes a stationary skyrmion structure and that in the second term drives the skyrmion into motion. The unit of the field gradient is $D^3/(J_{\mathrm{ex}}^2 M_s)$ with $M_s$ the saturation field. The torque is then given by
\begin{align}
\mathbf{\Gamma}  =-\mathbf{S}_0\times \left(\Delta_B^{(x)} r\cos\phi \hat{z}\right).
\end{align}
The coupling coefficient $F_{m,j}$ is
\begin{align}
F_{m,j}=-i \Delta_B^{(x)} \pi \delta_{|m|,1} \int dr r^2 L_{X;m,j} \sin\theta.
\end{align}
Here $F_{m, j}$ is nonzero only for $|m|=1$. In the rigid skyrmion approximation or $\omega\rightarrow0$, we have
\begin{align} 
F_{|m|=1,j=1}=\frac{i \pi\Delta_B^{(x)}}{\sqrt{8\pi\kappa}}\int dr r (\sin\theta)^2.
\end{align} 
The equation of motion is
\begin{align}\label{eq32}
\mathbf{v}(\omega)=-\frac{\Delta_B^{(x)}(\alpha  \hat{x}+\kappa  \hat{y})}{4 \left(\alpha ^2+\kappa ^2\right)}\int dr r (\sin\theta)^2.
\end{align}
For $\alpha\ll1$, the skyrmion moves almost perpendicular to the magnetic field gradient. There is additional velocity component antiparallel to the field gradient due to the damping. 

\begin{figure}[t]
\psfig{figure=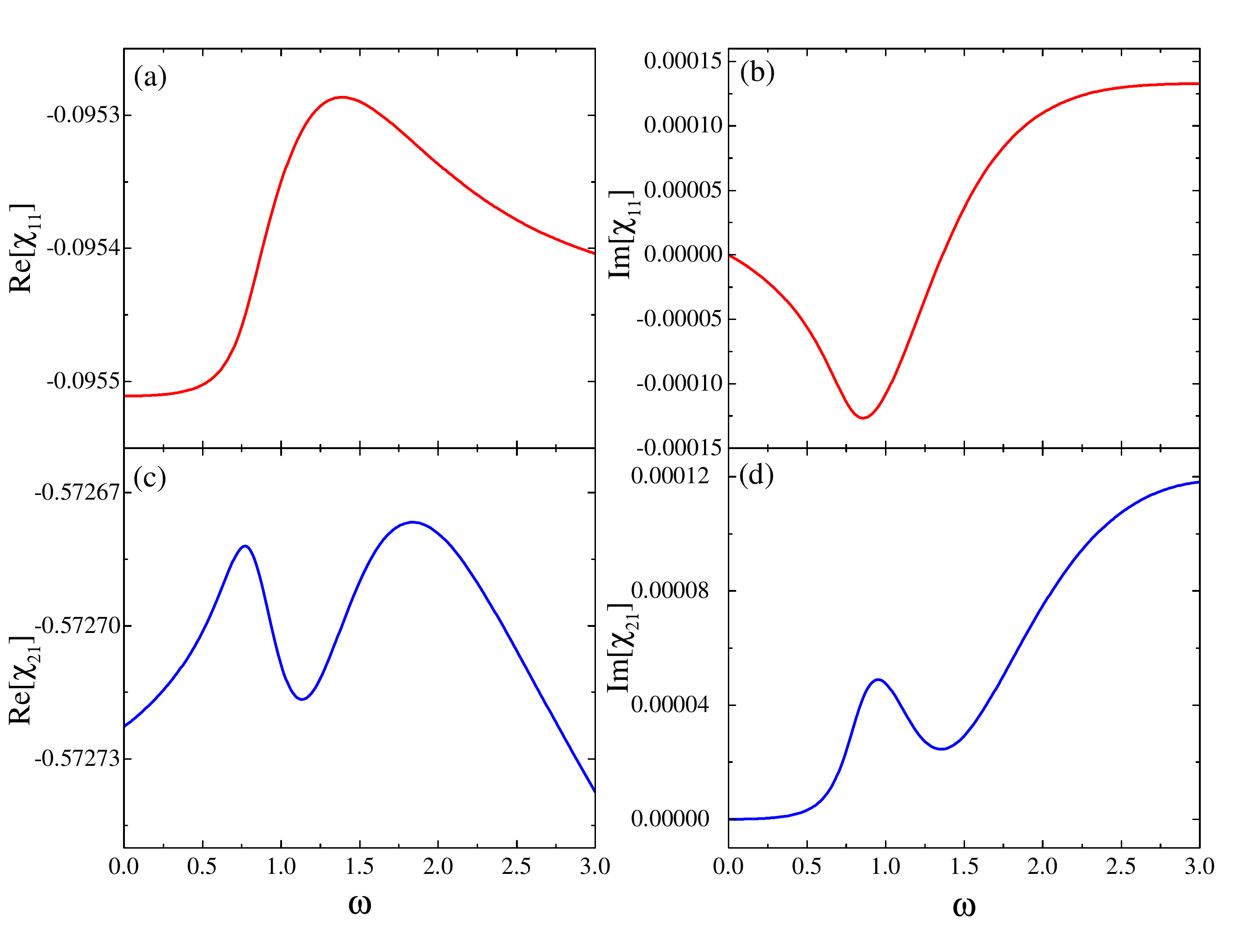,width=\columnwidth}
\caption{(color online) Real and imaginary part of $\chi_{11}$ and $\chi_{21}$ in the case of a skyrmion driven by a magnetic field gradient. Here $B=0.8$ and $\alpha=0.2$.
} \label{f2}
\end{figure}

The external drive in Eq. \eqref{eq25} is the field gradient $P_{\mu}=\Delta_B^{(\mu)}$. To include all the magnon modes, we calculate numerically the response matrix $\chi_{ij}$ and the results are shown in Fig. \ref{f2}. At $\omega=0$, $\chi_{ij}$ is real and the Thiele's equation of motion in Eq. \eqref{eq32} is recovered. We can see that $\mathrm{Im}[\chi_{ij}]$ is much smaller than $\mathrm{Re}[{\chi_{ij}}]$, because $\chi_{ij}$ is mainly contributed from the Goldstone modes. The phase shift near the magnon gap is $\Theta_{\parallel}\sim 10^{-3}$ and $\Theta_{\perp}\sim 10^{-4}$. The response is almost instantaneous and the inertia is weak. As shown in Fig. \ref{fa1}, the magnon modes associated with the Goldstone modes with $|m|=1$, $j=1$ are localized in the skyrmion, while the extended modes with $j>1$ only have small weight around the skyrmion. Therefore $|F_{|m|=1, j=1}|\gg |F_{|m|=1, j>1}|$ and $|\mathbf{W}_{|m|=1, j=1}|\gg |\mathbf{W}_{|m|=1, j>1}|$.

\begin{figure}[b]
\psfig{figure=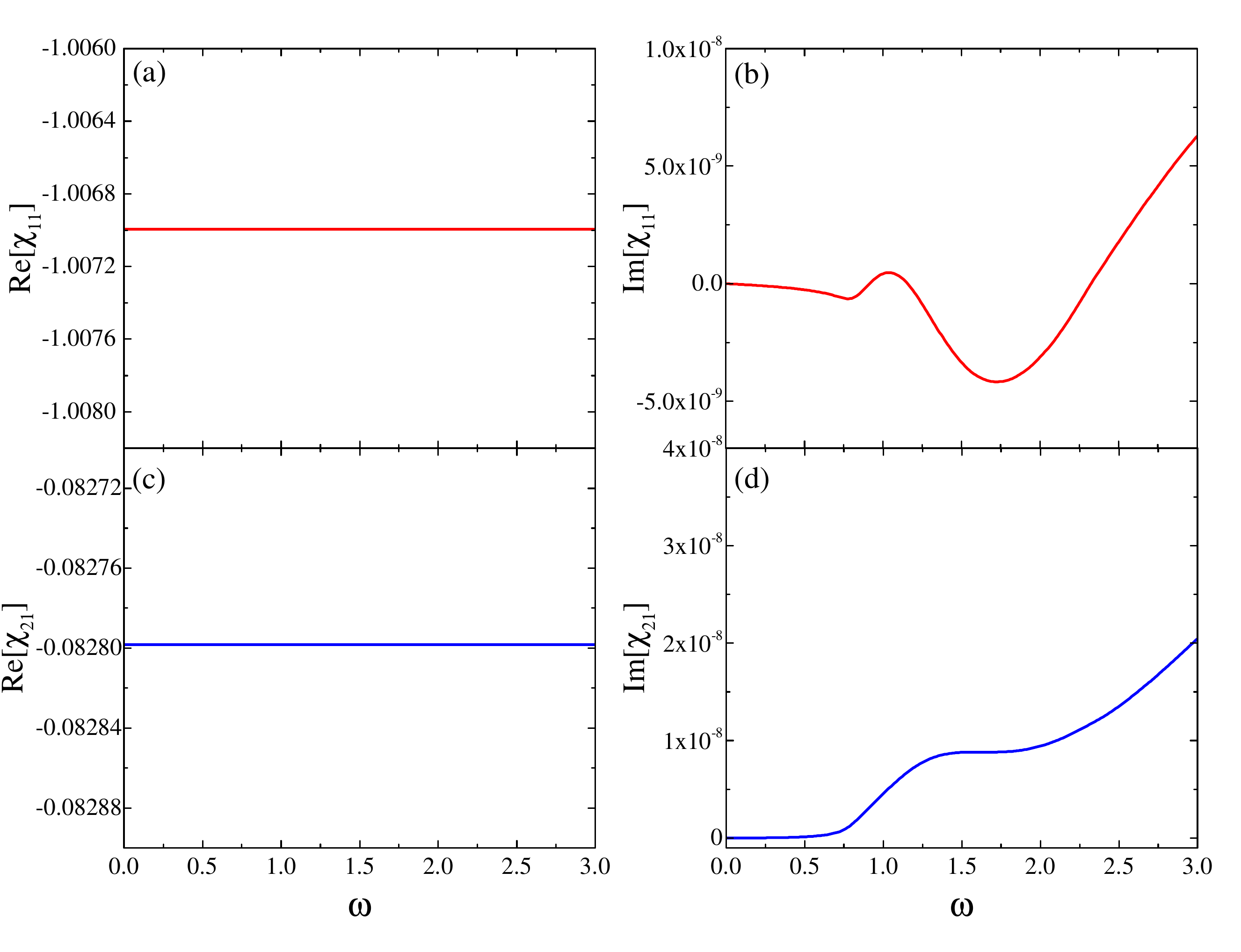,width=\columnwidth}
\caption{(color online) Real and imaginary part of $\chi_{11}$ and $\chi_{21}$ in the case of a skyrmion driven by a spin transfer torque. Here $B=0.8$, $\alpha=0.1$ and $\beta=0.2$.
} \label{f3}
\end{figure}

\subsection{Spin transfer torque}
Here we study the skyrmion motion driven by a spin transfer torque as shown in Fig. \ref{f1bb} (b). We inject an electric current into to a chiral magnet. The electric current is polarized automatically by the skyrmion spin texture because of the Hund's coupling, which results in a spin current. For a nonuniform spin texture, the polarization of spin current changes in space. To conserve magnetic moment, there is transfer of magnetic moment between the spin current and localized spin texture, which generates a torque acting on the localized moments. The spin transfer torque can be expressed as \cite{Slonczewski1996,Tatara2008}
\begin{align}
\mathbf{\Gamma}  = \left[ {{\partial _x}{\mathbf{S}_0} - \beta {\mathbf{S}_0} \times \left( {{\partial _x}{\mathbf{S}_0}} \right)} \right]{J_x}.
\end{align}
The current has unit of $2eD/(\hbar P_s)$ and is assumed to be along the $x$ direction, $\mathbf{J}=J_x\hat{x}$. The dimensionless parameter $P_s$ is the spin polarization factor.  In the first approximation, the direction of spin current polarization is always parallel to the local magnetization vector. This contribution, called adiabatic spin transfer torque, is described by the first term at the right-hand side.  The second term proportional to $\beta\ll 1$ describes the nonadiabatic and dissipative spin transfer torque, originating from the spatial mistracking of polarization of spin current and localized magnetization. \cite{PhysRevLett.93.127204} We then compute $F_{m,j}=-iJ_x(\tau_{m,j}^{\mathrm{STT}}+\beta \tau_{m,j}^{\mathrm{N}})$, with
\begin{align}
\tau_{m,j}^{\mathrm{STT}}=-\pi\int dr\left[ -i m L_{X;m,j}^*\sin(\theta)+r L_{Y;m,j}^*\partial_r\theta\right]\delta_{|m|,1}\delta_{j,1},
\end{align}
\begin{align}
\tau_{m,j}^{\mathrm{N}}=-\pi\int dr\left[ i m L_{Y;m,j}^*\sin(\theta)+r L_{X;m,j}^*\partial_r\theta\right]\delta_{|m|,1}.
\end{align}
The coupling coefficient  $F_{m,j}$ vanishes for $|m|\neq 1$. Furthermore, for the adiabatic spin transfer torque, $\tau_{m,j}^{\mathrm{STT}}$, the current couples with the translational mode $\partial_x\mathbf{S}_0$. Therefore $\tau_{m,j}^{\mathrm{STT}}$ is nonzero only for the translational mode with $j=1$. The inertia of a skyrmion arises from the contribution of the nonadiabatic spin transfer torque, $\tau_{m,j}^{\mathrm{N}}$, and damping.

For a rigid skyrmion valid when $\omega\rightarrow0$, we have
\begin{align}
F_{|m=1|,j=1}=J_x\left(-m\sqrt{2\pi\kappa}+i\sqrt{\frac{2\pi}{\kappa}}\beta\right),
\end{align}
The equation of motion becomes
\begin{align}\label{eq37}
\mathbf{v}(\omega)=\left[-\frac{\kappa^2+\alpha\beta}{\alpha^2+\kappa^2}\hat{x}+\frac{(\alpha-\beta)\kappa}{\alpha^2+\kappa^2}\hat{y}\right]J_x(\omega),
\end{align}
which reproduces the well known Thiele's equation of motion, Eq. \eqref{eq0}. It also describes the \emph{full} equation of motion when $\beta=0$ because the adiabatic spin transfer only couples with the translational mode. In real materials, $\alpha\ll 1$ and $\beta\ll 1$, the skyrmion moves almost antiparallel to current, but there is a transverse motion due to the damping. We can define a Hall angle for the skyrmion motion $\tan\theta_H\equiv |v_y|/|v_x|=|\alpha-\beta|\kappa/(\kappa^2+\alpha\beta)$. For $\alpha=\beta$, Eq. \eqref{eq15} is  Galilean invariant for a rigid skyrmion and the skyrmion moves exactly antiparallel to the current.

For $\beta>0$ and to deal with the skyrmion distortion, we compute numerically $\chi_{11}$ and $\chi_{21}$ at $B=0.8$ by taking the modes with $j>1$ into account, and the results are displayed in Fig. \ref{f3}. Here $\mathbf{P}$ in Eq. \eqref{eq25} is $\mathbf{P}=\mathbf{J}$. We have $\mathrm{Re}[\chi_{11}]\approx -({\kappa^2+\alpha\beta})/({\alpha^2+\kappa^2})$, $\mathrm{Re}[\chi_{21}]\approx {(\alpha-\beta)\kappa}/({\alpha^2+\kappa^2})$. The phase shift near the magnon gap is $\Theta_{\parallel}\sim 10^{-9}$ and $\Theta_{\perp}\sim 10^{-7}$. The contributions from other modes are negligible and the dynamics of a skyrmion is described by the Thiele's equation Eq. \eqref{eq37}. Therefore the response of a skyrmion to a spin transfer torque is instantaneous and the inertia of a skyrmion is negligible, compared to the viscous and the Magnus force. This justified the rigid skyrmion approximation used in Thiele's approach in this case.

\subsection{Spin Hall torque}

\begin{figure}[b]
\psfig{figure=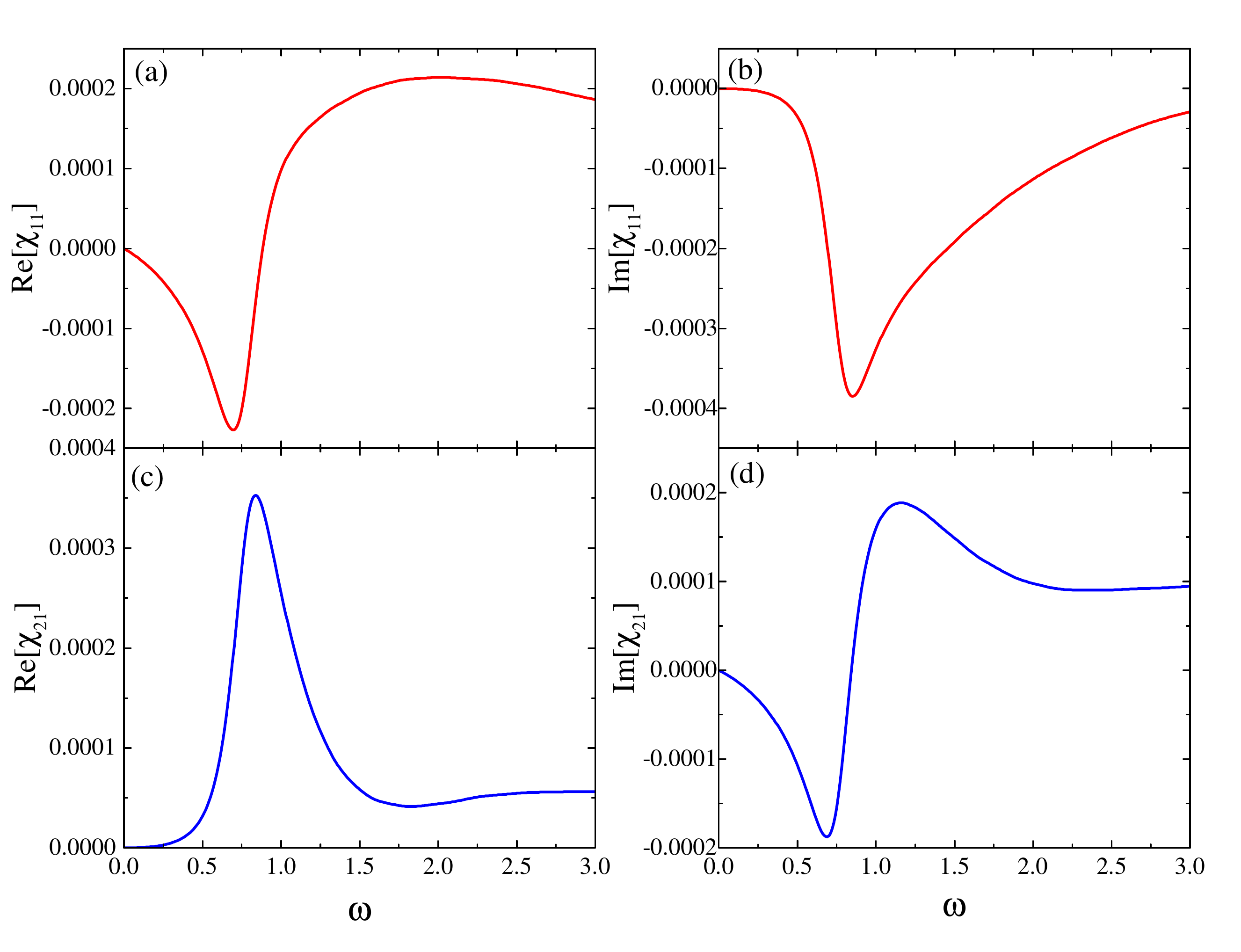,width=\columnwidth}
\caption{(color online) Real and imaginary part of $\chi_{11}$ and $\chi_{21}$ in the case of a skyrmion driven by a spin Hall torque. Here $B=0.8$ and $\alpha=0.2$.
} \label{f4}
\end{figure}

We consider a bilayer system with a chiral magnet atop of a heavy metal, such as Ta and Pd, as depicted in Fig. \ref{f1bb} (c). We then apply a current in the heavy metal. Because of the spin Hall effect, there is spin current normal to the interface, which generates a torque in the chiral magnet. This is called the  spin Hall torque and is given by \cite{emori_current-driven_2013,ryu_chiral_2013,PhysRevB.90.184427,liu_spin-torque_2012}
\begin{align}
\mathbf{\Gamma}={{\bf{S}}_0} \times \left( {{{\bf{S}}_0} \times \left( {\hat z \times {\bf{J}}} \right)} \right).
\end{align}
The current is in unit of $2 D^2 e d/(J_{\mathrm{ex}}\hbar\theta_{\mathrm{SH}})$ with $\theta_{\mathrm{SH}}$ being the spin Hall angle and $d$ the film thickness. For a current along the $x$ direction, $\mathbf{J}=J_x\hat{x}$, the coupling coefficient $F_{m,j}$ becomes
\begin{align}
F_{m,j}=-i\pi J_x\delta_{|m|,1} \int dr r (L_{Y;m,j}^* \cos\theta-i m L_{X;m,j}^*).
\end{align}
In the rigid skyrmion approximation or $\omega\rightarrow0$, 
\begin{align}
F_{|m|=1,j=1}=-i\int dr^2\left[\frac{{{{-i}m}{\partial _x}{\mathbf{S}_0} - {\partial _y}{\mathbf{S}_0}}}{{\sqrt 2 \sqrt {\int{{d}}{r^2}{{\left( {{\partial _x}{\mathbf{S}_0}} \right)}^2}} }}\right]\cdot\mathbf{\Gamma}.
\end{align}
We obtain $F_{|m|=1,j=1}=0$ because the skyrmion has rotational symmetry. Therefore the rigid skyrmion does not couple to the spin Hall torque when the damping is neglected. The torque couples to the skyrmion motion through excitation of the magnon modes in the continuum and the weak off diagonal damping term in Eq. \eqref{eqalphamj}. Therefore the spin Hall torque is less efficient in driving the skyrmion compared to that of spin transfer torque. Here we provide an estimate on the current density required in order to achieve the same velocity for the cases with spin transfer torque and spin Hall torque. For the spin transfer torque, the velocity is almost frequency independent for a given current; while for the spin Hall torque, the velocity is maximal at a given current when the frequency is near the magnon gap. Taking the optimal velocity for the spin Hall torque, to attain the same velocity, the required current density for both torques is about  $J_{\mathrm{SHE}}\theta_{\mathrm{SH}}J_{\mathrm{ex}}/J_{\mathrm{STT}} d D P_s\sim 10^3$, where $J_{\mathrm{SHE}}$ ($J_{\mathrm{STT}}$) is the current density in the case of spin Hall (transfer) torque. Therefore $J_{\mathrm{SHE}}$ is larger than $J_{\mathrm{STT}}$ by several orders of magnitude for materials with $\theta_{\mathrm{SH}}J_{\mathrm{ex}}/ d D P_s\sim 1$. This is consistent with the experimental observations that the skyrmion velocity at a given current in the case of spin transfer torque is much bigger than that in the case of spin Hall torque \cite{Schulz2012,Jiang17072015,jiang_direct_2017}, although different pinning strength in these systems may also partially account for the difference.

In the present case, $\mathbf{P}$ in Eq. \eqref{eq25} is $\mathbf{P}=\mathbf{J}$ .The numerical results of $\chi_{ij}$ are presented in Fig. \ref{f4}. Since $\chi_{21}\sim i \chi_{11} $, the trajectory is roughly a circle. One prominent feature is that $\chi_{ij}(\omega)$ develops a resonance-like feature around the magnon gap, $\omega\approx \omega_g$. This can be understood as follows. For a given angular momentum $|m|=1$, the density of state of magnon, $\rho(\omega)=1/\sqrt{\omega-\omega_g}$ ,  diverges at $\omega_g$. The equation of motion of a skyrmion by including all magnon excitations is
\begin{align}\label{eq40}
{\mathbf{R}}(\omega)=\sum_{|m|=1, j}\frac{F_{m,j}  \mathbf{W}_{m,j}}{\omega-\omega_{m,j}+i \alpha_{m, j}\omega} \nonumber\\
\approx F(\omega_g)  \mathbf{W}(\omega_g)\int_{\omega_g}^{+\infty} d\omega \rho(\omega)\frac{1}{\omega-\omega_{m,j}+i \alpha(\omega_g)\omega} \nonumber\\
=2F(\omega_g)  \mathbf{W}(\omega_g)\frac{1}{\sqrt{\omega-\omega_g+ i \alpha(\omega_g)\omega}},
\end{align}
where $F(\omega_g)$, $\mathbf{W}(\omega_g)$ and $\alpha(\omega_g)$ are the corresponding quantities evaluated at the magnon gap. From Eq. \eqref{eq40}, it is clear that $\chi_{ij}$ develops peaks around $\omega\approx \omega_g$, in agreement with the results in Fig. \ref{f4}.

\section{N\'{e}el skyrmion}
Skyrmion can also be stabilized at interfaces of heterostructure, where the inversion symmetry is broken explicitly \cite{Heinze2011,Fert2013,Romming2013}. The skyrmion at interface has helicity of $0$ and is called N\'{e}el skyrmion. The effective Hamiltonian supporting N\'{e}el skyrmions in two dimensions, $(x,\ y)$, can be written as \cite{thiaville_dynamics_2012,PhysRevB.90.184427}
\begin{equation}\label{Neeleq1}
\mathcal{H}_n= \frac{{{J_{\mathrm{ex}}}}}{2}\mathop \sum \limits_{\mu  = x,y} {\left( {{\partial _\mu }{\bf{n}}} \right)^2}+D[n_z(\nabla\cdot \mathbf{n})-(\mathbf{n}\cdot\nabla)n_z]-\mathbf{B}\cdot\mathbf{n}.
\end{equation}  
The Bloch skyrmion described by Eq. \eqref{eq1} can be obtained by a global rotation of spin associated with the N\'{e}el skyrmion along the magnetic field direction by $\pi/2$, i.e. $\mathbf{S}=\hat{O}_n \mathbf{n}$, with the spin rotation operator
\begin{equation}\label{Neeleq2}
\hat{O}_n= \left(
\begin{array}{ccc}
0 & -1 & 0 \\
1 & 0 & 0 \\
 0 & 0 & 1 \\
\end{array}
\right).
\end{equation}
One can verify that $\mathcal{H}_n$ can be obtained from $\mathcal{H}$ in Eq. \eqref{eq1} by the same global rotation of spins. Therefore the eigenfrequencies of the internal modes for the Bloch and N\'{e}el skyrmions are identical, and the eigenmodes for two skyrmion textures are related by a spin rotation.

One can also connect the equation of motion for the Bloch skyrmion to that of the N\'{e}el skyrmion by the same spin rotation. For a Bloch skyrmion driven by a magnetic field gradient with polarization perpendicular to the layer, spin transfer torque and spin Hall torque, the equation of motion for spins is
\begin{align}\label{Neeleq3}
{\partial _t}{\bf{S}} =  - {\bf{S}} \times {{\bf{H}}_{\mathrm{eff}}}(\mathbf{S}) + \alpha {\bf{S}} \times {\partial _t}{\bf{S}} + \left( {\mathbf{J}_{\mathrm{STT}}\cdot\nabla } \right){\bf{S}} \nonumber\\
- \beta \mathbf{S} \times \left( {\mathbf{J}_\mathrm{STT}\cdot\nabla } \right){\bf{S}} + {\bf{S}} \times \left[ {{\bf{S}} \times \left( {{\bf{\hat z}} \times {{\bf{J}}_{\mathrm{SHE}}}} \right)} \right].
\end{align}
After the spin rotation, the equation of motion for spins associated with a N\'{e}el skyrmion has the same form as Eq. \eqref{Neeleq3} if we replace the effective field by $\mathcal{H}_{\mathrm{eff}}(\mathbf{n})\equiv-\delta \mathcal{H}_n/\delta \mathbf{n}$ and $\mathbf{J}_{\mathrm{SHE}}\leftarrow \hat{O}_n^{-1} \mathbf{J}_{\mathrm{SHE}} $. Therefore the results derived for the Bloch skyrmions can be applied directly to the N\'{e}el skyrmions.

\section{Discussions and conclusions}\label{Sec4}
The current can generate a magnetic field gradient which affects the skyrmion trajectory. The magnetic field gradient depends on the sample geometry. In our linear approximation, we need to superpose the trajectory due to the current on the trajectory due to the induced field gradient. Meanwhile, the skyrmion motion distorts magnetic field outside of the magnetic plate, which contributes to the skyrmion kinetic energy. In the adiabatic approximation, a moving skyrmion induces an electric field $\nabla\times {\bf E}=(1/c)\dot{ {\bf B}}=-(\mathbf{v}\cdot\nabla) {\bf B}/c$ according to the Faraday's law. As a result its kinetic energy increases as $\int d^2r {\bf E}^2/(8\pi)=M_{\rm em}\dot{{\bf R}}^2$ which determines the electromagnetic part of the skyrmion mass tensor $M_{\rm em}$. Such an inertia effect has relativistic origin and is proportional to $(v/c)^2$. Since the maximal skyrmion velocity, $10^2\ \mathrm{m/s}$, \cite{szlin13skyrmion1} is much smaller than the light velocity, $10^8\ \mathrm{m/s}$, the correction to skyrmion dynamics due to the electric field outside the magnetic plate is negligible. 

Let us compare our results on the inertia of a skyrmion to those in literatures. The mass of a skyrmion bubble was calculated and measured in Refs. \onlinecite{Makhfudz2012,buttner_dynamics_2015} and they found that the mass term is important in order to reproduce the skyrmion trajectory. In the calculations \cite{Makhfudz2012}, the authors calculate the mass from the edge modes of the bubble. In experiments \cite{buttner_dynamics_2015}, the mass was obtained by fitting the experimentally measured trajectory determined from the spin polarization to equation of motion of a skyrmion as a particle with a mass term, assuming a hormonic potential for the skyrmion bubble. In Refs. \onlinecite{Makhfudz2012,buttner_dynamics_2015}, the skyrmion bubbles are stabilized by the dipolar interactions, and the magnon excitations are expected to be quite different from those of a skyrmion in chiral magnets. This may be the origin of the difference between our results the those in Refs. \onlinecite{Makhfudz2012,buttner_dynamics_2015}. This also suggests that the skyrmions in chiral magnets are more rigid and are advantageous for applications. The inertia of a skyrmion in chiral magnets was studied recently in Ref. \onlinecite{PhysRevB.90.174434}. They calculated numerically the trajectory of a skyrmion from the Landau-Lifshitz-Gilbert equation with a thermal noise, where the skyrmion diffuses in the sample. They then fit the trajectory to the equation of motion for a skyrmion as a particle according to Eq. \eqref{eq26}, from which they extracted the mass and gyrocoupling coefficient. They found that the inertia is important to describe the skyrmion diffusion. For a skyrmion driven by a spin transfer torque, they found the response of a skyrmion to current is almost instantaneous, consistent with our results. It is unclear the importance of nonlinearity in their calculations.  

The dynamics of skyrmions depends on its internal modes. The analysis here is applicable for skyrmions stabilized by the DM interaction, where the helicity of skyrmions is not a Goldstone mode. Skyrmions can also exist in inversion-symmetric magnets with competing interactions \cite{PhysRevLett.108.017206,leonov_multiply_2015,PhysRevB.93.064430,PhysRevB.93.184413}. Due to the preservation of the inversion symmetry, the helicity of skyrmion is also a Goldstone mode, in addition to the mode associated with translational motion. The dynamics can be very different, as has been demonstrated recently in Refs. \onlinecite{PhysRevB.93.064430,zhang_skyrmions_2017}. Skyrmions with higher topological charge can also be stabilized \cite{PhysRevB.95.094423}. The internal modes are different for skyrmions with higher topological charge, and the dynamics of these skyrmions requires a separate study.

To conclude, we have developed a linear theory to calculate the equation of motion of a skyrmion by taking all the magnon modes into account. We calculate the skyrmion velocity in response to external drives, such as a magnetic field gradient, spin transfer torque and spin Hall torque. The skyrmion dynamics is only governed by the magnon modes with an angular momentum $|m|=1$. The inertia of a skyrmion is contributed from the magnon continuum. For a skyrmion driven by a magnetic field gradient or spin transfer torque, the dynamics is dominated by the Goldstone modes corresponding to the translational motion of a skyrmion because the eigen function of the Goldstone modes have maximal weight around the skyrmion center, while the modes in the magnon continuum only has very little weight around the skyrmion. The response of a skyrmion is instantaneous and our calculations justify the rigid skyrmion approximation employed in the Thiele's collective coordinate approach. In the case of a spin Hall torque, the skyrmion motion couples to the torque through the magnons in the continuum and the Gilbert damping. Since the magnon density of state diverges around the magnon gap, there are resonances in the response functions around the magnon gap. The trajectory of a skyrmion is an ellipse for a skyrmion driven by a spin Hall torque at finite frequencies. The inertia of a skyrmion can be quantified in experiments by measuring the phase shift between the external drive and velocity. For applications, it is desirable for skyrmion to respond to an external drive without delay or retardation even at high frequencies. This can be achieved by driving a skyrmion with a spin transfer torque.  Our results establish the connection between the skyrmion dynamics and its magnon spectrum, and shed new light on the skyrmion dynamics. 

\begin{acknowledgments}
The author is indebted to Lev N. Bulaevskii, Avadh Saxena, Cristian D. Batista and Satoru Hayami  for helpful discussions. This work was carried out under the auspices of the U.S. DOE Contract No. DE-AC52-06NA25396 through the LDRD program.
\end{acknowledgments}

\appendix
\section{Numerical method for the eigenmodes analysis}

\begin{figure}[t]
\psfig{figure=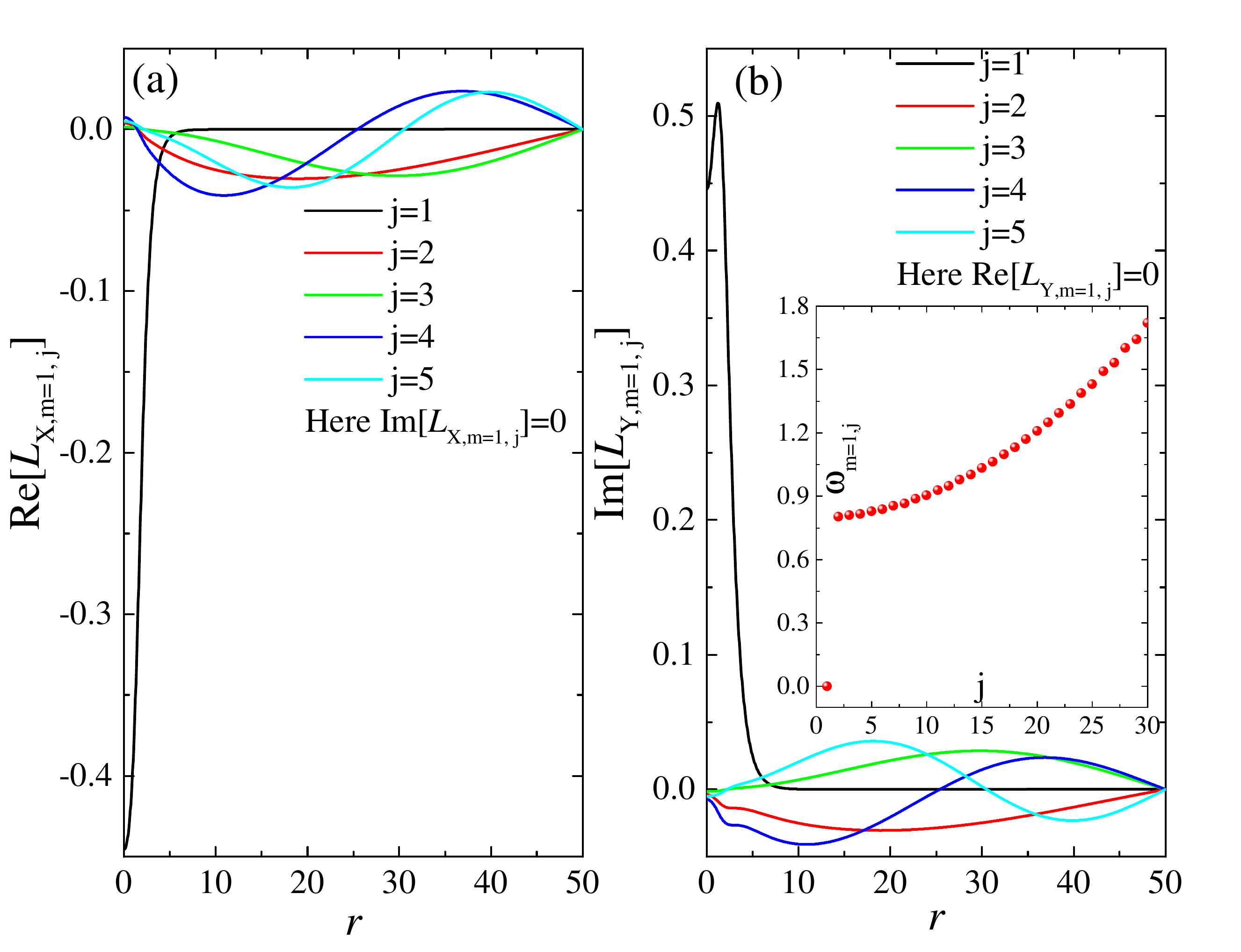,width=\columnwidth}
\caption{(color online) Five lowest magnon wave functions with $m=1$ in the rotated frame. $L_{X;m=1, j}$ is real and $L_{Y;m=1, j}$ is imaginary. Inset is the eigenfrequency. Here $B=0.8$.
} \label{fa1}
\end{figure}

\begin{figure}[b]
\psfig{figure=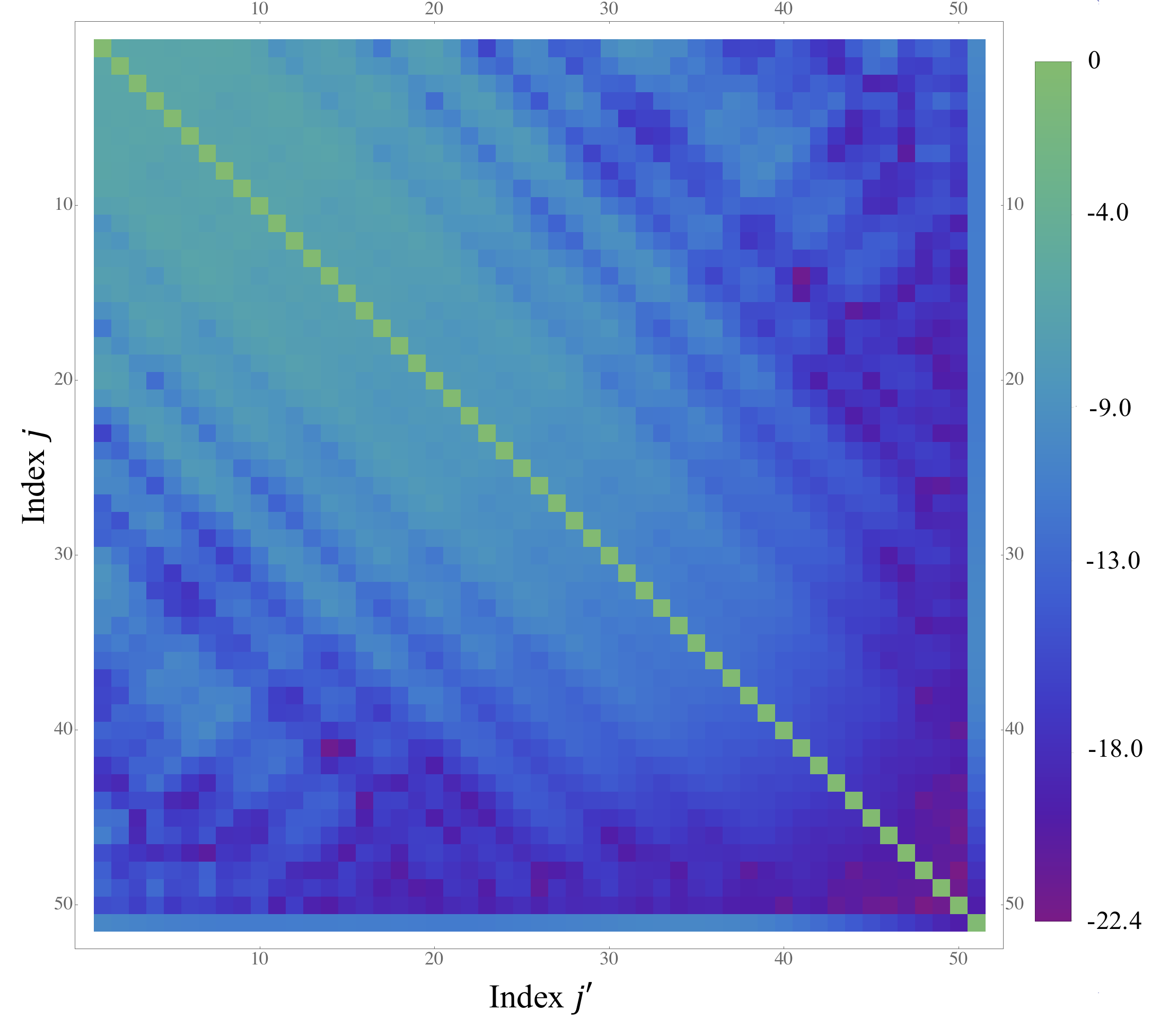,width=\columnwidth}
\caption{(color online) Results of $\ln(|\alpha_{m=-1, j, j'}|/\alpha)$ for $1\le j,\ j'\le 51$ in the frequency region $0\le\omega_{m, j}\le 11$ obtained at $B=0.8$. The green line in the diagonal direction is the dominant component $\ln(|\alpha_{m=-1, j, j}|/\alpha)$.
} \label{fa2}
\end{figure}

We present the details about the numerical evaluation of the eigenmodes in Eq. \eqref{eq13}. We introduce an orthogonal complete basis to expand $\eta_m$ in this basis. In the polar coordinate, a natural choice is the Bessel functions of the first kind. To determine the proper Bessel functions, we first check $\hat{\eta}_m$ in the $r\ll 1$ limit in Eq. \eqref{eq13}, which can be solved analytically
\begin{equation}
\hat{\eta}_m=\left( {\begin{array}{*{20}{c}}
{{c_ - }{r^{\left| {m - 1} \right|}}}\\
{{c_ + }{r^{\left| {m + 1} \right|}}}
\end{array}} \right).
\end{equation}
From the asymptotic behavior of $\hat{\eta}_m$, we choose the following basis
\begin{align}
|{p_{m,i}}\rangle  = \frac{{\sqrt 2 }}{{{R_c}{J_m}\left( {{k_{m - 1,i}}} \right)}}{J_{m - 1}}\left( {{k_{m - 1,i}}\frac{r}{{{R_c}}}} \right)\exp \left( {{\rm{i}}m\phi } \right)\left( {\begin{array}{*{20}{c}}
1\\
0
\end{array}} \right),
\end{align}
\begin{align}
|{h_{m,i}}\rangle  = \frac{{\sqrt 2 }}{{{R_c}{J_{m + 2}}\left( {{k_{m + 1,i}}} \right)}}{J_{m + 1}}\left( {{k_{m + 1,i}}\frac{r}{{{R_c}}}} \right)\exp \left( {{\rm{i}}m\phi } \right)\left( {\begin{array}{*{20}{c}}
0\\
1
\end{array}} \right),
\end{align}
where we have used the box normalization with $R_c$ the radius of the box and $k_{m,i}$ is the $i$-th zero of the Bessel function $J_m(k)$. Then the matrix element of $\mathcal{H}_\psi$ is
\begin{align}
\hat{\mathcal{H}}_{11;ij}^{(m)}=\langle p_{m,i} | \mathcal{H}_\psi |{p_{m,j}}\rangle, \ \  \hat{\mathcal{H}}_{12;ij}^{(m)}=\langle p_{m,i} | \mathcal{H}_\psi |{h_{m,j}}\rangle, \nonumber\\
\hat{\mathcal{H}}_{21;ij}^{(m)}=\langle h_{m,i} | \mathcal{H}_\psi |{p_{m,j}}\rangle, \ \  \hat{\mathcal{H}}_{22;ij}^{(m)}=\langle h_{m,i} | \mathcal{H}_\psi |{h_{m,j}}\rangle.
\end{align}
By diagonalizing the matrix $\sigma_z \mathcal{H}_\psi$, we obtain the eigenfrequencies and eigenmodes. We take $R_c=50$ and truncate the Bessel series at $i_{\mathrm{max}}=100$. The results for the eigen modes and eigen frequencies with $m=1$ in the rotated frame are shown in Fig.~\ref{fa1}. There is only one bound state with $\omega_{m=1, j=1}=0$ corresponding to the translational motion of  skyrmion. The wave function of the bound state has large weight around the skyrmion center while the magnon modes in the continuum have little weight around the skyrmion.

\section{Evaluation of the damping matrix $\alpha_{m, j, j'}$}
We calculate numerically $\alpha_{m=-1, j, j'}$ for $1\le j,\ j'\le 51$ in the frequency region $0\le\omega_{m, j}\le 11$. The results of $\ln(|\alpha_{m=-1, j, j'}|/\alpha)$ are displayed in Fig. \ref{fa2}. The diagonal elements $|\alpha_{m, j, j}|$ are of several orders of magnitude larger than the off diagonal elements $|\alpha_{m, j, j'}|$, indicating that the overlap between different magnon modes in Eq. \eqref{eqalphamj} is negligible. This justifies the approximation used in the previous section, where only the diagonal elements are taken into account,  $\alpha_{m, j, j'}\approx \alpha_{m, j}\delta_{j, j'}$. 

\bibliography{reference}

\begin{thebibliography}{52}%
\makeatletter
\providecommand \@ifxundefined [1]{%
 \@ifx{#1\undefined}
}%
\providecommand \@ifnum [1]{%
 \ifnum #1\expandafter \@firstoftwo
 \else \expandafter \@secondoftwo
 \fi
}%
\providecommand \@ifx [1]{%
 \ifx #1\expandafter \@firstoftwo
 \else \expandafter \@secondoftwo
 \fi
}%
\providecommand \natexlab [1]{#1}%
\providecommand \enquote  [1]{``#1''}%
\providecommand \bibnamefont  [1]{#1}%
\providecommand \bibfnamefont [1]{#1}%
\providecommand \citenamefont [1]{#1}%
\providecommand \href@noop [0]{\@secondoftwo}%
\providecommand \href [0]{\begingroup \@sanitize@url \@href}%
\providecommand \@href[1]{\@@startlink{#1}\@@href}%
\providecommand \@@href[1]{\endgroup#1\@@endlink}%
\providecommand \@sanitize@url [0]{\catcode `\\12\catcode `\$12\catcode
  `\&12\catcode `\#12\catcode `\^12\catcode `\_12\catcode `\%12\relax}%
\providecommand \@@startlink[1]{}%
\providecommand \@@endlink[0]{}%
\providecommand \url  [0]{\begingroup\@sanitize@url \@url }%
\providecommand \@url [1]{\endgroup\@href {#1}{\urlprefix }}%
\providecommand \urlprefix  [0]{URL }%
\providecommand \Eprint [0]{\href }%
\providecommand \doibase [0]{http://dx.doi.org/}%
\providecommand \selectlanguage [0]{\@gobble}%
\providecommand \bibinfo  [0]{\@secondoftwo}%
\providecommand \bibfield  [0]{\@secondoftwo}%
\providecommand \translation [1]{[#1]}%
\providecommand \BibitemOpen [0]{}%
\providecommand \bibitemStop [0]{}%
\providecommand \bibitemNoStop [0]{.\EOS\space}%
\providecommand \EOS [0]{\spacefactor3000\relax}%
\providecommand \BibitemShut  [1]{\csname bibitem#1\endcsname}%
\let\auto@bib@innerbib\@empty
\bibitem [{\citenamefont {Skyrme}(1961)}]{Skyrme61}%
  \BibitemOpen
  \bibfield  {author} {\bibinfo {author} {\bibfnamefont {T.~H.~R.}\
  \bibnamefont {Skyrme}},\ }\bibfield  {title} {\enquote {\bibinfo {title} {A
  non-linear field theory},}\ }\href {\doibase 10.1098/rspa.1961.0018}
  {\bibfield  {journal} {\bibinfo  {journal} {Proc. R. Soc. A}\ }\textbf
  {\bibinfo {volume} {260}},\ \bibinfo {pages} {127--138} (\bibinfo {year}
  {1961})}\BibitemShut {NoStop}%
\bibitem [{\citenamefont {Skyrme}(1962)}]{Skyrme1962556}%
  \BibitemOpen
  \bibfield  {author} {\bibinfo {author} {\bibfnamefont {T.H.R.}\ \bibnamefont
  {Skyrme}},\ }\bibfield  {title} {\enquote {\bibinfo {title} {A unified field
  theory of mesons and baryons},}\ }\href {\doibase
  http://dx.doi.org/10.1016/0029-5582(62)90775-7} {\bibfield  {journal}
  {\bibinfo  {journal} {Nuclear Physics}\ }\textbf {\bibinfo {volume} {31}},\
  \bibinfo {pages} {556 -- 569} (\bibinfo {year} {1962})}\BibitemShut {NoStop}%
\bibitem [{\citenamefont {Bogdanov}\ and\ \citenamefont
  {Yablonskii}(1989)}]{Bogdanov89}%
  \BibitemOpen
  \bibfield  {author} {\bibinfo {author} {\bibfnamefont {A.~N.}\ \bibnamefont
  {Bogdanov}}\ and\ \bibinfo {author} {\bibfnamefont {D.~A.}\ \bibnamefont
  {Yablonskii}},\ }\bibfield  {title} {\enquote {\bibinfo {title}
  {Thermodynamically stable ``vortices" in magnetically ordered crystals: The
  mixed state of magnets},}\ }\href@noop {} {\bibfield  {journal} {\bibinfo
  {journal} {Sov. Phys. JETP}\ }\textbf {\bibinfo {volume} {68}},\ \bibinfo
  {pages} {101} (\bibinfo {year} {1989})}\BibitemShut {NoStop}%
\bibitem [{\citenamefont {M\"{u}hlbauer}\ \emph {et~al.}(2009)\citenamefont
  {M\"{u}hlbauer}, \citenamefont {Binz}, \citenamefont {Jonietz}, \citenamefont
  {Pfleiderer}, \citenamefont {Rosch}, \citenamefont {Neubauer}, \citenamefont
  {Georgii},\ and\ \citenamefont {B\"{o}ni}}]{Muhlbauer2009}%
  \BibitemOpen
  \bibfield  {author} {\bibinfo {author} {\bibfnamefont {S.}~\bibnamefont
  {M\"{u}hlbauer}}, \bibinfo {author} {\bibfnamefont {B.}~\bibnamefont {Binz}},
  \bibinfo {author} {\bibfnamefont {F.}~\bibnamefont {Jonietz}}, \bibinfo
  {author} {\bibfnamefont {C.}~\bibnamefont {Pfleiderer}}, \bibinfo {author}
  {\bibfnamefont {A.}~\bibnamefont {Rosch}}, \bibinfo {author} {\bibfnamefont
  {A.}~\bibnamefont {Neubauer}}, \bibinfo {author} {\bibfnamefont
  {R.}~\bibnamefont {Georgii}}, \ and\ \bibinfo {author} {\bibfnamefont
  {P.}~\bibnamefont {B\"{o}ni}},\ }\bibfield  {title} {\enquote {\bibinfo
  {title} {Skyrmion lattice in a chiral magnet},}\ }\href {\doibase
  10.1126/science.1166767} {\bibfield  {journal} {\bibinfo  {journal}
  {Science}\ }\textbf {\bibinfo {volume} {323}},\ \bibinfo {pages} {915}
  (\bibinfo {year} {2009})}\BibitemShut {NoStop}%
\bibitem [{\citenamefont {Yu}\ \emph {et~al.}(2010)\citenamefont {Yu},
  \citenamefont {Onose}, \citenamefont {Kanazawa}, \citenamefont {Park},
  \citenamefont {Han}, \citenamefont {Matsui}, \citenamefont {Nagaosa},\ and\
  \citenamefont {Tokura}}]{Yu2010a}%
  \BibitemOpen
  \bibfield  {author} {\bibinfo {author} {\bibfnamefont {X.~Z.}\ \bibnamefont
  {Yu}}, \bibinfo {author} {\bibfnamefont {Y.}~\bibnamefont {Onose}}, \bibinfo
  {author} {\bibfnamefont {N.}~\bibnamefont {Kanazawa}}, \bibinfo {author}
  {\bibfnamefont {J.~H.}\ \bibnamefont {Park}}, \bibinfo {author}
  {\bibfnamefont {J.~H.}\ \bibnamefont {Han}}, \bibinfo {author} {\bibfnamefont
  {Y.}~\bibnamefont {Matsui}}, \bibinfo {author} {\bibfnamefont
  {N.}~\bibnamefont {Nagaosa}}, \ and\ \bibinfo {author} {\bibfnamefont
  {Y.}~\bibnamefont {Tokura}},\ }\bibfield  {title} {\enquote {\bibinfo {title}
  {Real-space observation of a two-dimensional skyrmion crystal},}\ }\href
  {\doibase 10.1038/nature09124} {\bibfield  {journal} {\bibinfo  {journal}
  {Nature}\ }\textbf {\bibinfo {volume} {465}},\ \bibinfo {pages} {901}
  (\bibinfo {year} {2010})}\BibitemShut {NoStop}%
\bibitem [{\citenamefont {Jonietz}\ \emph {et~al.}(2010)\citenamefont
  {Jonietz}, \citenamefont {M\"uhlbauer}, \citenamefont {Pfleiderer},
  \citenamefont {Neubauer}, \citenamefont {M\"unzer}, \citenamefont {Bauer},
  \citenamefont {Adams}, \citenamefont {Georgii}, \citenamefont {B\"oni},
  \citenamefont {Duine}, \citenamefont {Everschor}, \citenamefont {Garst},\
  and\ \citenamefont {Rosch}}]{Jonietz2010}%
  \BibitemOpen
  \bibfield  {author} {\bibinfo {author} {\bibfnamefont {F.}~\bibnamefont
  {Jonietz}}, \bibinfo {author} {\bibfnamefont {S.}~\bibnamefont
  {M\"uhlbauer}}, \bibinfo {author} {\bibfnamefont {C.}~\bibnamefont
  {Pfleiderer}}, \bibinfo {author} {\bibfnamefont {A.}~\bibnamefont
  {Neubauer}}, \bibinfo {author} {\bibfnamefont {W.}~\bibnamefont {M\"unzer}},
  \bibinfo {author} {\bibfnamefont {A.}~\bibnamefont {Bauer}}, \bibinfo
  {author} {\bibfnamefont {T.}~\bibnamefont {Adams}}, \bibinfo {author}
  {\bibfnamefont {R.}~\bibnamefont {Georgii}}, \bibinfo {author} {\bibfnamefont
  {P.}~\bibnamefont {B\"oni}}, \bibinfo {author} {\bibfnamefont {R.~A.}\
  \bibnamefont {Duine}}, \bibinfo {author} {\bibfnamefont {K.}~\bibnamefont
  {Everschor}}, \bibinfo {author} {\bibfnamefont {M.}~\bibnamefont {Garst}}, \
  and\ \bibinfo {author} {\bibfnamefont {A.}~\bibnamefont {Rosch}},\ }\bibfield
   {title} {\enquote {\bibinfo {title} {Spin transfer torques in {MnSi} at
  ultralow current densities},}\ }\href {\doibase 10.1126/science.1195709}
  {\bibfield  {journal} {\bibinfo  {journal} {Science}\ }\textbf {\bibinfo
  {volume} {330}},\ \bibinfo {pages} {1648} (\bibinfo {year}
  {2010})}\BibitemShut {NoStop}%
\bibitem [{\citenamefont {Yu}\ \emph {et~al.}(2012)\citenamefont {Yu},
  \citenamefont {Kanazawa}, \citenamefont {Zhang}, \citenamefont {Nagai},
  \citenamefont {Hara}, \citenamefont {Kimoto}, \citenamefont {Matsui},
  \citenamefont {Onose},\ and\ \citenamefont {Tokura}}]{Yu2012}%
  \BibitemOpen
  \bibfield  {author} {\bibinfo {author} {\bibfnamefont {X.~Z.}\ \bibnamefont
  {Yu}}, \bibinfo {author} {\bibfnamefont {N.}~\bibnamefont {Kanazawa}},
  \bibinfo {author} {\bibfnamefont {W.~Z.}\ \bibnamefont {Zhang}}, \bibinfo
  {author} {\bibfnamefont {T.}~\bibnamefont {Nagai}}, \bibinfo {author}
  {\bibfnamefont {T.}~\bibnamefont {Hara}}, \bibinfo {author} {\bibfnamefont
  {K.}~\bibnamefont {Kimoto}}, \bibinfo {author} {\bibfnamefont
  {Y.}~\bibnamefont {Matsui}}, \bibinfo {author} {\bibfnamefont
  {Y.}~\bibnamefont {Onose}}, \ and\ \bibinfo {author} {\bibfnamefont
  {Y.}~\bibnamefont {Tokura}},\ }\bibfield  {title} {\enquote {\bibinfo {title}
  {Skyrmion flow near room temperature in an ultralow current density},}\
  }\href {\doibase 10.1038/ncomms1990} {\bibfield  {journal} {\bibinfo
  {journal} {Nat. Commun.}\ }\textbf {\bibinfo {volume} {3}},\ \bibinfo {pages}
  {988} (\bibinfo {year} {2012})}\BibitemShut {NoStop}%
\bibitem [{\citenamefont {Schulz}\ \emph {et~al.}(2012)\citenamefont {Schulz},
  \citenamefont {Ritz}, \citenamefont {Bauer}, \citenamefont {Halder},
  \citenamefont {Wagner}, \citenamefont {Franz}, \citenamefont {Pfleiderer},
  \citenamefont {Everschor}, \citenamefont {Garst},\ and\ \citenamefont
  {Rosch}}]{Schulz2012}%
  \BibitemOpen
  \bibfield  {author} {\bibinfo {author} {\bibfnamefont {T.}~\bibnamefont
  {Schulz}}, \bibinfo {author} {\bibfnamefont {R.}~\bibnamefont {Ritz}},
  \bibinfo {author} {\bibfnamefont {A.}~\bibnamefont {Bauer}}, \bibinfo
  {author} {\bibfnamefont {M.}~\bibnamefont {Halder}}, \bibinfo {author}
  {\bibfnamefont {M.}~\bibnamefont {Wagner}}, \bibinfo {author} {\bibfnamefont
  {C.}~\bibnamefont {Franz}}, \bibinfo {author} {\bibfnamefont
  {C.}~\bibnamefont {Pfleiderer}}, \bibinfo {author} {\bibfnamefont
  {K.}~\bibnamefont {Everschor}}, \bibinfo {author} {\bibfnamefont
  {M.}~\bibnamefont {Garst}}, \ and\ \bibinfo {author} {\bibfnamefont
  {A.}~\bibnamefont {Rosch}},\ }\bibfield  {title} {\enquote {\bibinfo {title}
  {Emergent electrodynamics of skyrmions in a chiral magnet},}\ }\href
  {\doibase 10.1038/nphys2231} {\bibfield  {journal} {\bibinfo  {journal} {Nat.
  Phys.}\ }\textbf {\bibinfo {volume} {8}},\ \bibinfo {pages} {301} (\bibinfo
  {year} {2012})}\BibitemShut {NoStop}%
\bibitem [{\citenamefont {White}\ \emph {et~al.}(2012)\citenamefont {White},
  \citenamefont {Levatic}, \citenamefont {Omrani}, \citenamefont {Egetenmeyer},
  \citenamefont {Prsa}, \citenamefont {Zivkovic}, \citenamefont {Gavilano},
  \citenamefont {Kohlbrecher}, \citenamefont {Bartkowiak}, \citenamefont
  {Berger},\ and\ \citenamefont {Ronnow}}]{White2012}%
  \BibitemOpen
  \bibfield  {author} {\bibinfo {author} {\bibfnamefont {J.~S.}\ \bibnamefont
  {White}}, \bibinfo {author} {\bibfnamefont {I.}~\bibnamefont {Levatic}},
  \bibinfo {author} {\bibfnamefont {A.~A.}\ \bibnamefont {Omrani}}, \bibinfo
  {author} {\bibfnamefont {N.}~\bibnamefont {Egetenmeyer}}, \bibinfo {author}
  {\bibfnamefont {K.}~\bibnamefont {Prsa}}, \bibinfo {author} {\bibfnamefont
  {I.}~\bibnamefont {Zivkovic}}, \bibinfo {author} {\bibfnamefont {J.~L.}\
  \bibnamefont {Gavilano}}, \bibinfo {author} {\bibfnamefont {J.}~\bibnamefont
  {Kohlbrecher}}, \bibinfo {author} {\bibfnamefont {M.}~\bibnamefont
  {Bartkowiak}}, \bibinfo {author} {\bibfnamefont {H.}~\bibnamefont {Berger}},
  \ and\ \bibinfo {author} {\bibfnamefont {H.~M.}\ \bibnamefont {Ronnow}},\
  }\bibfield  {title} {\enquote {\bibinfo {title} {Electric field control of
  the skyrmion lattice in {$\mathrm{Cu_2OSeO_3}$}},}\ }\href {\doibase
  10.1088/0953-8984/24/43/432201} {\bibfield  {journal} {\bibinfo  {journal}
  {J. Phys.: Condens. Matter}\ }\textbf {\bibinfo {volume} {24}},\ \bibinfo
  {pages} {432201} (\bibinfo {year} {2012})}\BibitemShut {NoStop}%
\bibitem [{\citenamefont {White}\ \emph {et~al.}(2014)\citenamefont {White},
  \citenamefont {Pr\ifmmode~\check{s}\else \v{s}\fi{}a}, \citenamefont {Huang},
  \citenamefont {Omrani}, \citenamefont {\ifmmode \check{Z}\else
  \v{Z}\fi{}ivkovi\ifmmode~\acute{c}\else \'{c}\fi{}}, \citenamefont
  {Bartkowiak}, \citenamefont {Berger}, \citenamefont {Magrez}, \citenamefont
  {Gavilano}, \citenamefont {Nagy}, \citenamefont {Zang},\ and\ \citenamefont
  {R\o{}nnow}}]{PhysRevLett.113.107203}%
  \BibitemOpen
  \bibfield  {author} {\bibinfo {author} {\bibfnamefont {J.~S.}\ \bibnamefont
  {White}}, \bibinfo {author} {\bibfnamefont {K.}~\bibnamefont
  {Pr\ifmmode~\check{s}\else \v{s}\fi{}a}}, \bibinfo {author} {\bibfnamefont
  {P.}~\bibnamefont {Huang}}, \bibinfo {author} {\bibfnamefont {A.~A.}\
  \bibnamefont {Omrani}}, \bibinfo {author} {\bibfnamefont {I.}~\bibnamefont
  {\ifmmode \check{Z}\else \v{Z}\fi{}ivkovi\ifmmode~\acute{c}\else
  \'{c}\fi{}}}, \bibinfo {author} {\bibfnamefont {M.}~\bibnamefont
  {Bartkowiak}}, \bibinfo {author} {\bibfnamefont {H.}~\bibnamefont {Berger}},
  \bibinfo {author} {\bibfnamefont {A.}~\bibnamefont {Magrez}}, \bibinfo
  {author} {\bibfnamefont {J.~L.}\ \bibnamefont {Gavilano}}, \bibinfo {author}
  {\bibfnamefont {G.}~\bibnamefont {Nagy}}, \bibinfo {author} {\bibfnamefont
  {J.}~\bibnamefont {Zang}}, \ and\ \bibinfo {author} {\bibfnamefont {H.~M.}\
  \bibnamefont {R\o{}nnow}},\ }\bibfield  {title} {\enquote {\bibinfo {title}
  {Electric-field-induced skyrmion distortion and giant lattice rotation in the
  magnetoelectric insulator {$\mathrm{Cu_2OSeO_3}$}},}\ }\href {\doibase
  10.1103/PhysRevLett.113.107203} {\bibfield  {journal} {\bibinfo  {journal}
  {Phys. Rev. Lett.}\ }\textbf {\bibinfo {volume} {113}},\ \bibinfo {pages}
  {107203} (\bibinfo {year} {2014})}\BibitemShut {NoStop}%
\bibitem [{\citenamefont {Kong}\ and\ \citenamefont {Zang}(2013)}]{Kong2013}%
  \BibitemOpen
  \bibfield  {author} {\bibinfo {author} {\bibfnamefont {Lingyao}\ \bibnamefont
  {Kong}}\ and\ \bibinfo {author} {\bibfnamefont {Jiadong}\ \bibnamefont
  {Zang}},\ }\bibfield  {title} {\enquote {\bibinfo {title} {Dynamics of an
  insulating skyrmion under a temperature gradient},}\ }\href {\doibase
  10.1103/PhysRevLett.111.067203} {\bibfield  {journal} {\bibinfo  {journal}
  {Phys. Rev. Lett.}\ }\textbf {\bibinfo {volume} {111}},\ \bibinfo {pages}
  {067203} (\bibinfo {year} {2013})}\BibitemShut {NoStop}%
\bibitem [{\citenamefont {Lin}\ \emph {et~al.}(2014{\natexlab{a}})\citenamefont
  {Lin}, \citenamefont {Batista}, \citenamefont {Reichhardt},\ and\
  \citenamefont {Saxena}}]{Lin2014PRL}%
  \BibitemOpen
  \bibfield  {author} {\bibinfo {author} {\bibfnamefont {Shi-Zeng}\
  \bibnamefont {Lin}}, \bibinfo {author} {\bibfnamefont {Cristian~D.}\
  \bibnamefont {Batista}}, \bibinfo {author} {\bibfnamefont {Charles}\
  \bibnamefont {Reichhardt}}, \ and\ \bibinfo {author} {\bibfnamefont {Avadh}\
  \bibnamefont {Saxena}},\ }\bibfield  {title} {\enquote {\bibinfo {title}
  {{ac} current generation in chiral magnetic insulators and skyrmion motion
  induced by the spin {Seebeck} effect},}\ }\href {\doibase
  10.1103/PhysRevLett.112.187203} {\bibfield  {journal} {\bibinfo  {journal}
  {Phys. Rev. Lett.}\ }\textbf {\bibinfo {volume} {112}},\ \bibinfo {pages}
  {187203} (\bibinfo {year} {2014}{\natexlab{a}})}\BibitemShut {NoStop}%
\bibitem [{\citenamefont {Mochizuki}\ \emph {et~al.}(2014)\citenamefont
  {Mochizuki}, \citenamefont {Yu}, \citenamefont {Seki}, \citenamefont
  {Kanazawa}, \citenamefont {Koshibae}, \citenamefont {Zang}, \citenamefont
  {Mostovoy}, \citenamefont {Tokura},\ and\ \citenamefont
  {Nagaosa}}]{Mochizuki2014}%
  \BibitemOpen
  \bibfield  {author} {\bibinfo {author} {\bibfnamefont {M.}~\bibnamefont
  {Mochizuki}}, \bibinfo {author} {\bibfnamefont {X.~Z.}\ \bibnamefont {Yu}},
  \bibinfo {author} {\bibfnamefont {S.}~\bibnamefont {Seki}}, \bibinfo {author}
  {\bibfnamefont {N.}~\bibnamefont {Kanazawa}}, \bibinfo {author}
  {\bibfnamefont {W.}~\bibnamefont {Koshibae}}, \bibinfo {author}
  {\bibfnamefont {J.}~\bibnamefont {Zang}}, \bibinfo {author} {\bibfnamefont
  {M.}~\bibnamefont {Mostovoy}}, \bibinfo {author} {\bibfnamefont
  {Y.}~\bibnamefont {Tokura}}, \ and\ \bibinfo {author} {\bibfnamefont
  {N.}~\bibnamefont {Nagaosa}},\ }\bibfield  {title} {\enquote {\bibinfo
  {title} {Thermally driven ratchet motion of a skyrmion microcrystal and
  topological magnon {Hall} effect},}\ }\href {\doibase 10.1038/nmat3862}
  {\bibfield  {journal} {\bibinfo  {journal} {Nature Materials}\ }\textbf
  {\bibinfo {volume} {13}},\ \bibinfo {pages} {241} (\bibinfo {year}
  {2014})}\BibitemShut {NoStop}%
\bibitem [{\citenamefont {Sch\"utte}\ and\ \citenamefont
  {Garst}(2014)}]{PhysRevB.90.094423}%
  \BibitemOpen
  \bibfield  {author} {\bibinfo {author} {\bibfnamefont {Christoph}\
  \bibnamefont {Sch\"utte}}\ and\ \bibinfo {author} {\bibfnamefont {Markus}\
  \bibnamefont {Garst}},\ }\bibfield  {title} {\enquote {\bibinfo {title}
  {Magnon-skyrmion scattering in chiral magnets},}\ }\href {\doibase
  10.1103/PhysRevB.90.094423} {\bibfield  {journal} {\bibinfo  {journal} {Phys.
  Rev. B}\ }\textbf {\bibinfo {volume} {90}},\ \bibinfo {pages} {094423}
  (\bibinfo {year} {2014})}\BibitemShut {NoStop}%
\bibitem [{\citenamefont {Fert}\ \emph {et~al.}(2013)\citenamefont {Fert},
  \citenamefont {Cros},\ and\ \citenamefont {Sampaio}}]{Fert2013}%
  \BibitemOpen
  \bibfield  {author} {\bibinfo {author} {\bibfnamefont {Albert}\ \bibnamefont
  {Fert}}, \bibinfo {author} {\bibfnamefont {Vincent}\ \bibnamefont {Cros}}, \
  and\ \bibinfo {author} {\bibfnamefont {J.}~\bibnamefont {Sampaio}},\
  }\bibfield  {title} {\enquote {\bibinfo {title} {Skyrmions on the track},}\
  }\href {\doibase 10.1038/nnano.2013.29} {\bibfield  {journal} {\bibinfo
  {journal} {Nat. Nanotechnol.}\ }\textbf {\bibinfo {volume} {8}},\ \bibinfo
  {pages} {152--156} (\bibinfo {year} {2013})}\BibitemShut {NoStop}%
\bibitem [{\citenamefont {Nagaosa}\ and\ \citenamefont
  {Tokura}(2013)}]{nagaosa_topological_2013}%
  \BibitemOpen
  \bibfield  {author} {\bibinfo {author} {\bibfnamefont {Naoto}\ \bibnamefont
  {Nagaosa}}\ and\ \bibinfo {author} {\bibfnamefont {Yoshinori}\ \bibnamefont
  {Tokura}},\ }\bibfield  {title} {\enquote {\bibinfo {title} {Topological
  properties and dynamics of magnetic skyrmions},}\ }\href {\doibase
  10.1038/nnano.2013.243} {\bibfield  {journal} {\bibinfo  {journal} {Nature
  Nanotechnology}\ }\textbf {\bibinfo {volume} {8}},\ \bibinfo {pages}
  {899--911} (\bibinfo {year} {2013})}\BibitemShut {NoStop}%
\bibitem [{\citenamefont {Thiele}(1973)}]{Thiele72}%
  \BibitemOpen
  \bibfield  {author} {\bibinfo {author} {\bibfnamefont {A.~A.}\ \bibnamefont
  {Thiele}},\ }\bibfield  {title} {\enquote {\bibinfo {title} {Steady-state
  motion of magnetic domains},}\ }\href {\doibase 10.1103/PhysRevLett.30.230}
  {\bibfield  {journal} {\bibinfo  {journal} {Phys. Rev. Lett.}\ }\textbf
  {\bibinfo {volume} {30}},\ \bibinfo {pages} {230--233} (\bibinfo {year}
  {1973})}\BibitemShut {NoStop}%
\bibitem [{\citenamefont {Iwasaki}\ \emph {et~al.}(2013)\citenamefont
  {Iwasaki}, \citenamefont {Mochizuki},\ and\ \citenamefont
  {Nagaosa}}]{Iwasaki2013}%
  \BibitemOpen
  \bibfield  {author} {\bibinfo {author} {\bibfnamefont {Junichi}\ \bibnamefont
  {Iwasaki}}, \bibinfo {author} {\bibfnamefont {Masahito}\ \bibnamefont
  {Mochizuki}}, \ and\ \bibinfo {author} {\bibfnamefont {Naoto}\ \bibnamefont
  {Nagaosa}},\ }\bibfield  {title} {\enquote {\bibinfo {title} {Universal
  current-velocity relation of skyrmion motion in chiral magnets},}\ }\href
  {\doibase 10.1038/ncomms2442} {\bibfield  {journal} {\bibinfo  {journal}
  {Nat. Commun.}\ }\textbf {\bibinfo {volume} {4}},\ \bibinfo {pages} {1463}
  (\bibinfo {year} {2013})}\BibitemShut {NoStop}%
\bibitem [{\citenamefont {Lin}\ \emph {et~al.}(2013{\natexlab{a}})\citenamefont
  {Lin}, \citenamefont {Reichhardt}, \citenamefont {Batista},\ and\
  \citenamefont {Saxena}}]{szlin13skyrmion1}%
  \BibitemOpen
  \bibfield  {author} {\bibinfo {author} {\bibfnamefont {Shi-Zeng}\
  \bibnamefont {Lin}}, \bibinfo {author} {\bibfnamefont {Charles}\ \bibnamefont
  {Reichhardt}}, \bibinfo {author} {\bibfnamefont {Cristian~D.}\ \bibnamefont
  {Batista}}, \ and\ \bibinfo {author} {\bibfnamefont {Avadh}\ \bibnamefont
  {Saxena}},\ }\bibfield  {title} {\enquote {\bibinfo {title} {Driven skyrmions
  and dynamical transitions in chiral magnets},}\ }\href {\doibase
  10.1103/PhysRevLett.110.207202} {\bibfield  {journal} {\bibinfo  {journal}
  {Phys. Rev. Lett.}\ }\textbf {\bibinfo {volume} {110}},\ \bibinfo {pages}
  {207202} (\bibinfo {year} {2013}{\natexlab{a}})}\BibitemShut {NoStop}%
\bibitem [{\citenamefont {Lin}\ \emph {et~al.}(2013{\natexlab{b}})\citenamefont
  {Lin}, \citenamefont {Reichhardt}, \citenamefont {Batista},\ and\
  \citenamefont {Saxena}}]{szlin13skyrmion2}%
  \BibitemOpen
  \bibfield  {author} {\bibinfo {author} {\bibfnamefont {Shi-Zeng}\
  \bibnamefont {Lin}}, \bibinfo {author} {\bibfnamefont {Charles}\ \bibnamefont
  {Reichhardt}}, \bibinfo {author} {\bibfnamefont {Cristian~D.}\ \bibnamefont
  {Batista}}, \ and\ \bibinfo {author} {\bibfnamefont {Avadh}\ \bibnamefont
  {Saxena}},\ }\bibfield  {title} {\enquote {\bibinfo {title} {Particle model
  for skyrmions in metallic chiral magnets: Dynamics, pinning, and creep},}\
  }\href {\doibase 10.1103/PhysRevB.87.214419} {\bibfield  {journal} {\bibinfo
  {journal} {Phys. Rev. B}\ }\textbf {\bibinfo {volume} {87}},\ \bibinfo
  {pages} {214419} (\bibinfo {year} {2013}{\natexlab{b}})}\BibitemShut
  {NoStop}%
\bibitem [{\citenamefont {Lin}\ \emph {et~al.}(2014{\natexlab{b}})\citenamefont
  {Lin}, \citenamefont {Batista},\ and\ \citenamefont
  {Saxena}}]{Lin_internal_2014}%
  \BibitemOpen
  \bibfield  {author} {\bibinfo {author} {\bibfnamefont {Shi-Zeng}\
  \bibnamefont {Lin}}, \bibinfo {author} {\bibfnamefont {Cristian~D.}\
  \bibnamefont {Batista}}, \ and\ \bibinfo {author} {\bibfnamefont {Avadh}\
  \bibnamefont {Saxena}},\ }\bibfield  {title} {\enquote {\bibinfo {title}
  {Internal modes of a skyrmion in the ferromagnetic state of chiral
  magnets},}\ }\href {\doibase 10.1103/PhysRevB.89.024415} {\bibfield
  {journal} {\bibinfo  {journal} {Phys. Rev. B}\ }\textbf {\bibinfo {volume}
  {89}},\ \bibinfo {pages} {024415} (\bibinfo {year}
  {2014}{\natexlab{b}})}\BibitemShut {NoStop}%
\bibitem [{\citenamefont {Everschor}\ \emph {et~al.}(2011)\citenamefont
  {Everschor}, \citenamefont {Garst}, \citenamefont {Duine},\ and\
  \citenamefont {Rosch}}]{Everschor11}%
  \BibitemOpen
  \bibfield  {author} {\bibinfo {author} {\bibfnamefont {Karin}\ \bibnamefont
  {Everschor}}, \bibinfo {author} {\bibfnamefont {Markus}\ \bibnamefont
  {Garst}}, \bibinfo {author} {\bibfnamefont {R.~A.}\ \bibnamefont {Duine}}, \
  and\ \bibinfo {author} {\bibfnamefont {Achim}\ \bibnamefont {Rosch}},\
  }\bibfield  {title} {\enquote {\bibinfo {title} {Current-induced rotational
  torques in the skyrmion lattice phase of chiral magnets},}\ }\href {\doibase
  10.1103/PhysRevB.84.064401} {\bibfield  {journal} {\bibinfo  {journal} {Phys.
  Rev. B}\ }\textbf {\bibinfo {volume} {84}},\ \bibinfo {pages} {064401}
  (\bibinfo {year} {2011})}\BibitemShut {NoStop}%
\bibitem [{\citenamefont {Everschor}\ \emph {et~al.}(2012)\citenamefont
  {Everschor}, \citenamefont {Garst}, \citenamefont {Binz}, \citenamefont
  {Jonietz}, \citenamefont {M\"uhlbauer}, \citenamefont {Pfleiderer},\ and\
  \citenamefont {Rosch}}]{Everschor12}%
  \BibitemOpen
  \bibfield  {author} {\bibinfo {author} {\bibfnamefont {Karin}\ \bibnamefont
  {Everschor}}, \bibinfo {author} {\bibfnamefont {Markus}\ \bibnamefont
  {Garst}}, \bibinfo {author} {\bibfnamefont {Benedikt}\ \bibnamefont {Binz}},
  \bibinfo {author} {\bibfnamefont {Florian}\ \bibnamefont {Jonietz}}, \bibinfo
  {author} {\bibfnamefont {Sebastian}\ \bibnamefont {M\"uhlbauer}}, \bibinfo
  {author} {\bibfnamefont {Christian}\ \bibnamefont {Pfleiderer}}, \ and\
  \bibinfo {author} {\bibfnamefont {Achim}\ \bibnamefont {Rosch}},\ }\bibfield
  {title} {\enquote {\bibinfo {title} {Rotating skyrmion lattices by spin
  torques and field or temperature gradients},}\ }\href {\doibase
  10.1103/PhysRevB.86.054432} {\bibfield  {journal} {\bibinfo  {journal} {Phys.
  Rev. B}\ }\textbf {\bibinfo {volume} {86}},\ \bibinfo {pages} {054432}
  (\bibinfo {year} {2012})}\BibitemShut {NoStop}%
\bibitem [{\citenamefont {Abanov}\ and\ \citenamefont
  {Pokrovsky}(1998)}]{PhysRevB.58.R8889}%
  \BibitemOpen
  \bibfield  {author} {\bibinfo {author} {\bibfnamefont {Ar.}\ \bibnamefont
  {Abanov}}\ and\ \bibinfo {author} {\bibfnamefont {V.~L.}\ \bibnamefont
  {Pokrovsky}},\ }\bibfield  {title} {\enquote {\bibinfo {title} {Skyrmion in a
  real magnetic film},}\ }\href {\doibase 10.1103/PhysRevB.58.R8889} {\bibfield
   {journal} {\bibinfo  {journal} {Phys. Rev. B}\ }\textbf {\bibinfo {volume}
  {58}},\ \bibinfo {pages} {R8889--R8892} (\bibinfo {year} {1998})}\BibitemShut
  {NoStop}%
\bibitem [{\citenamefont {V\"olkel}\ \emph {et~al.}(1994)\citenamefont
  {V\"olkel}, \citenamefont {Wysin}, \citenamefont {Mertens}, \citenamefont
  {Bishop},\ and\ \citenamefont {Schnitzer}}]{PhysRevB.50.12711}%
  \BibitemOpen
  \bibfield  {author} {\bibinfo {author} {\bibfnamefont {A.~R.}\ \bibnamefont
  {V\"olkel}}, \bibinfo {author} {\bibfnamefont {G.~M.}\ \bibnamefont {Wysin}},
  \bibinfo {author} {\bibfnamefont {F.~G.}\ \bibnamefont {Mertens}}, \bibinfo
  {author} {\bibfnamefont {A.~R.}\ \bibnamefont {Bishop}}, \ and\ \bibinfo
  {author} {\bibfnamefont {H.~J.}\ \bibnamefont {Schnitzer}},\ }\bibfield
  {title} {\enquote {\bibinfo {title} {Collective-variable approach to the
  dynamics of nonlinear magnetic excitations with application to vortices},}\
  }\href {\doibase 10.1103/PhysRevB.50.12711} {\bibfield  {journal} {\bibinfo
  {journal} {Phys. Rev. B}\ }\textbf {\bibinfo {volume} {50}},\ \bibinfo
  {pages} {12711--12720} (\bibinfo {year} {1994})}\BibitemShut {NoStop}%
\bibitem [{\citenamefont {Sch\"utte}\ \emph {et~al.}(2014)\citenamefont
  {Sch\"utte}, \citenamefont {Iwasaki}, \citenamefont {Rosch},\ and\
  \citenamefont {Nagaosa}}]{PhysRevB.90.174434}%
  \BibitemOpen
  \bibfield  {author} {\bibinfo {author} {\bibfnamefont {Christoph}\
  \bibnamefont {Sch\"utte}}, \bibinfo {author} {\bibfnamefont {Junichi}\
  \bibnamefont {Iwasaki}}, \bibinfo {author} {\bibfnamefont {Achim}\
  \bibnamefont {Rosch}}, \ and\ \bibinfo {author} {\bibfnamefont {Naoto}\
  \bibnamefont {Nagaosa}},\ }\bibfield  {title} {\enquote {\bibinfo {title}
  {Inertia, diffusion, and dynamics of a driven skyrmion},}\ }\href {\doibase
  10.1103/PhysRevB.90.174434} {\bibfield  {journal} {\bibinfo  {journal} {Phys.
  Rev. B}\ }\textbf {\bibinfo {volume} {90}},\ \bibinfo {pages} {174434}
  (\bibinfo {year} {2014})}\BibitemShut {NoStop}%
\bibitem [{\citenamefont {Büttner}\ \emph {et~al.}(2015)\citenamefont
  {Büttner}, \citenamefont {Moutafis}, \citenamefont {Schneider},
  \citenamefont {Krüger}, \citenamefont {Günther}, \citenamefont {Geilhufe},
  \citenamefont {Schmising}, \citenamefont {Mohanty}, \citenamefont {Pfau},
  \citenamefont {Schaffert}, \citenamefont {Bisig}, \citenamefont {Foerster},
  \citenamefont {Schulz}, \citenamefont {Vaz}, \citenamefont {Franken},
  \citenamefont {Swagten}, \citenamefont {Kläui},\ and\ \citenamefont
  {Eisebitt}}]{buttner_dynamics_2015}%
  \BibitemOpen
  \bibfield  {author} {\bibinfo {author} {\bibfnamefont {Felix}\ \bibnamefont
  {Büttner}}, \bibinfo {author} {\bibfnamefont {C.}~\bibnamefont {Moutafis}},
  \bibinfo {author} {\bibfnamefont {M.}~\bibnamefont {Schneider}}, \bibinfo
  {author} {\bibfnamefont {B.}~\bibnamefont {Krüger}}, \bibinfo {author}
  {\bibfnamefont {C.~M.}\ \bibnamefont {Günther}}, \bibinfo {author}
  {\bibfnamefont {J.}~\bibnamefont {Geilhufe}}, \bibinfo {author}
  {\bibfnamefont {C.~v~Korff}\ \bibnamefont {Schmising}}, \bibinfo {author}
  {\bibfnamefont {J.}~\bibnamefont {Mohanty}}, \bibinfo {author} {\bibfnamefont
  {B.}~\bibnamefont {Pfau}}, \bibinfo {author} {\bibfnamefont {S.}~\bibnamefont
  {Schaffert}}, \bibinfo {author} {\bibfnamefont {A.}~\bibnamefont {Bisig}},
  \bibinfo {author} {\bibfnamefont {M.}~\bibnamefont {Foerster}}, \bibinfo
  {author} {\bibfnamefont {T.}~\bibnamefont {Schulz}}, \bibinfo {author}
  {\bibfnamefont {C.~a.~F.}\ \bibnamefont {Vaz}}, \bibinfo {author}
  {\bibfnamefont {J.~H.}\ \bibnamefont {Franken}}, \bibinfo {author}
  {\bibfnamefont {H.~J.~M.}\ \bibnamefont {Swagten}}, \bibinfo {author}
  {\bibfnamefont {M.}~\bibnamefont {Kläui}}, \ and\ \bibinfo {author}
  {\bibfnamefont {S.}~\bibnamefont {Eisebitt}},\ }\bibfield  {title} {\enquote
  {\bibinfo {title} {Dynamics and inertia of skyrmionic spin structures},}\
  }\href {\doibase 10.1038/nphys3234} {\bibfield  {journal} {\bibinfo
  {journal} {Nature Physics}\ }\textbf {\bibinfo {volume} {11}},\ \bibinfo
  {pages} {225--228} (\bibinfo {year} {2015})}\BibitemShut {NoStop}%
\bibitem [{\citenamefont {Dzyaloshinsky}(1958)}]{Dzyaloshinsky1958}%
  \BibitemOpen
  \bibfield  {author} {\bibinfo {author} {\bibfnamefont {I.}~\bibnamefont
  {Dzyaloshinsky}},\ }\bibfield  {title} {\enquote {\bibinfo {title} {A
  thermodynamic theory of weak ferromagnetism of antiferromagnetics},}\ }\href
  {\doibase 10.1016/0022-3697(58)90076-3} {\bibfield  {journal} {\bibinfo
  {journal} {J. Phys. Chem. Solids}\ }\textbf {\bibinfo {volume} {4}},\
  \bibinfo {pages} {241} (\bibinfo {year} {1958})}\BibitemShut {NoStop}%
\bibitem [{\citenamefont {Moriya}(1960{\natexlab{a}})}]{Moriya60}%
  \BibitemOpen
  \bibfield  {author} {\bibinfo {author} {\bibfnamefont {T\^oru}\ \bibnamefont
  {Moriya}},\ }\bibfield  {title} {\enquote {\bibinfo {title} {Anisotropic
  superexchange interaction and weak ferromagnetism},}\ }\href {\doibase
  10.1103/PhysRev.120.91} {\bibfield  {journal} {\bibinfo  {journal} {Phys.
  Rev.}\ }\textbf {\bibinfo {volume} {120}},\ \bibinfo {pages} {91} (\bibinfo
  {year} {1960}{\natexlab{a}})}\BibitemShut {NoStop}%
\bibitem [{\citenamefont {Moriya}(1960{\natexlab{b}})}]{Moriya60b}%
  \BibitemOpen
  \bibfield  {author} {\bibinfo {author} {\bibfnamefont {T\^oru}\ \bibnamefont
  {Moriya}},\ }\bibfield  {title} {\enquote {\bibinfo {title} {New mechanism of
  anisotropic superexchange interaction},}\ }\href {\doibase
  10.1103/PhysRevLett.4.228} {\bibfield  {journal} {\bibinfo  {journal} {Phys.
  Rev. Lett.}\ }\textbf {\bibinfo {volume} {4}},\ \bibinfo {pages} {228--230}
  (\bibinfo {year} {1960}{\natexlab{b}})}\BibitemShut {NoStop}%
\bibitem [{\citenamefont {Bogdanov}\ and\ \citenamefont
  {Hubert}(1994)}]{Bogdanov94}%
  \BibitemOpen
  \bibfield  {author} {\bibinfo {author} {\bibfnamefont {A.}~\bibnamefont
  {Bogdanov}}\ and\ \bibinfo {author} {\bibfnamefont {A.}~\bibnamefont
  {Hubert}},\ }\bibfield  {title} {\enquote {\bibinfo {title}
  {Thermodynamically stable magnetic vortex states in magnetic crystals},}\
  }\href {\doibase http://dx.doi.org/10.1016/0304-8853(94)90046-9} {\bibfield
  {journal} {\bibinfo  {journal} {J. Magn. Magn. Mater.}\ }\textbf {\bibinfo
  {volume} {138}},\ \bibinfo {pages} {255 -- 269} (\bibinfo {year}
  {1994})}\BibitemShut {NoStop}%
\bibitem [{\citenamefont {Moon}\ \emph {et~al.}(2015)\citenamefont {Moon},
  \citenamefont {Kim}, \citenamefont {Yoo}, \citenamefont {Je}, \citenamefont
  {Chun}, \citenamefont {Kim}, \citenamefont {Min}, \citenamefont {Hwang},\
  and\ \citenamefont {Choe}}]{moon_magnetic_2015}%
  \BibitemOpen
  \bibfield  {author} {\bibinfo {author} {\bibfnamefont {Kyoung-Woong}\
  \bibnamefont {Moon}}, \bibinfo {author} {\bibfnamefont {Duck-Ho}\
  \bibnamefont {Kim}}, \bibinfo {author} {\bibfnamefont {Sang-Cheol}\
  \bibnamefont {Yoo}}, \bibinfo {author} {\bibfnamefont {Soong-Geun}\
  \bibnamefont {Je}}, \bibinfo {author} {\bibfnamefont {Byong~Sun}\
  \bibnamefont {Chun}}, \bibinfo {author} {\bibfnamefont {Wondong}\
  \bibnamefont {Kim}}, \bibinfo {author} {\bibfnamefont {Byoung-Chul}\
  \bibnamefont {Min}}, \bibinfo {author} {\bibfnamefont {Chanyong}\
  \bibnamefont {Hwang}}, \ and\ \bibinfo {author} {\bibfnamefont {Sug-Bong}\
  \bibnamefont {Choe}},\ }\bibfield  {title} {\enquote {\bibinfo {title}
  {Magnetic bubblecade memory based on chiral domain walls},}\ }\href {\doibase
  10.1038/srep09166} {\bibfield  {journal} {\bibinfo  {journal} {Scientific
  Reports}\ }\textbf {\bibinfo {volume} {5}},\ \bibinfo {pages} {9166}
  (\bibinfo {year} {2015})}\BibitemShut {NoStop}%
\bibitem [{\citenamefont {Moon}\ \emph {et~al.}(2016)\citenamefont {Moon},
  \citenamefont {Kim}, \citenamefont {Je}, \citenamefont {Chun}, \citenamefont
  {Kim}, \citenamefont {Qiu}, \citenamefont {Choe},\ and\ \citenamefont
  {Hwang}}]{moon_skyrmion_2016}%
  \BibitemOpen
  \bibfield  {author} {\bibinfo {author} {\bibfnamefont {Kyoung-Woong}\
  \bibnamefont {Moon}}, \bibinfo {author} {\bibfnamefont {Duck-Ho}\
  \bibnamefont {Kim}}, \bibinfo {author} {\bibfnamefont {Soong-Geun}\
  \bibnamefont {Je}}, \bibinfo {author} {\bibfnamefont {Byong~Sun}\
  \bibnamefont {Chun}}, \bibinfo {author} {\bibfnamefont {Wondong}\
  \bibnamefont {Kim}}, \bibinfo {author} {\bibfnamefont {Z.~Q.}\ \bibnamefont
  {Qiu}}, \bibinfo {author} {\bibfnamefont {Sug-Bong}\ \bibnamefont {Choe}}, \
  and\ \bibinfo {author} {\bibfnamefont {Chanyong}\ \bibnamefont {Hwang}},\
  }\bibfield  {title} {\enquote {\bibinfo {title} {Skyrmion motion driven by
  oscillating magnetic field},}\ }\href {\doibase 10.1038/srep20360} {\bibfield
   {journal} {\bibinfo  {journal} {Scientific Reports}\ }\textbf {\bibinfo
  {volume} {6}},\ \bibinfo {pages} {20360} (\bibinfo {year}
  {2016})}\BibitemShut {NoStop}%
\bibitem [{\citenamefont {Slonczewski}(1996)}]{Slonczewski1996}%
  \BibitemOpen
  \bibfield  {author} {\bibinfo {author} {\bibfnamefont {J.C.}\ \bibnamefont
  {Slonczewski}},\ }\bibfield  {title} {\enquote {\bibinfo {title}
  {Current-driven excitation of magnetic multilayers},}\ }\href {\doibase
  10.1016/0304-8853(96)00062-5} {\bibfield  {journal} {\bibinfo  {journal}
  {Journal of Magnetism and Magnetic Materials}\ }\textbf {\bibinfo {volume}
  {159}},\ \bibinfo {pages} {L1--L7} (\bibinfo {year} {1996})}\BibitemShut
  {NoStop}%
\bibitem [{\citenamefont {Tatara}\ \emph {et~al.}(2008)\citenamefont {Tatara},
  \citenamefont {Kohno},\ and\ \citenamefont {Shibata}}]{Tatara2008}%
  \BibitemOpen
  \bibfield  {author} {\bibinfo {author} {\bibfnamefont {Gen}\ \bibnamefont
  {Tatara}}, \bibinfo {author} {\bibfnamefont {Hiroshi}\ \bibnamefont {Kohno}},
  \ and\ \bibinfo {author} {\bibfnamefont {Junya}\ \bibnamefont {Shibata}},\
  }\bibfield  {title} {\enquote {\bibinfo {title} {Microscopic approach to
  current-driven domain wall dynamics},}\ }\href {\doibase
  10.1016/j.physrep.2008.07.003} {\bibfield  {journal} {\bibinfo  {journal}
  {Phys. Rep.}\ }\textbf {\bibinfo {volume} {468}},\ \bibinfo {pages} {213}
  (\bibinfo {year} {2008})}\BibitemShut {NoStop}%
\bibitem [{\citenamefont {Zhang}\ and\ \citenamefont
  {Li}(2004)}]{PhysRevLett.93.127204}%
  \BibitemOpen
  \bibfield  {author} {\bibinfo {author} {\bibfnamefont {S.}~\bibnamefont
  {Zhang}}\ and\ \bibinfo {author} {\bibfnamefont {Z.}~\bibnamefont {Li}},\
  }\bibfield  {title} {\enquote {\bibinfo {title} {Roles of nonequilibrium
  conduction electrons on the magnetization dynamics of ferromagnets},}\ }\href
  {\doibase 10.1103/PhysRevLett.93.127204} {\bibfield  {journal} {\bibinfo
  {journal} {Phys. Rev. Lett.}\ }\textbf {\bibinfo {volume} {93}},\ \bibinfo
  {pages} {127204} (\bibinfo {year} {2004})}\BibitemShut {NoStop}%
\bibitem [{\citenamefont {Emori}\ \emph {et~al.}(2013)\citenamefont {Emori},
  \citenamefont {Bauer}, \citenamefont {Ahn}, \citenamefont {Martinez},\ and\
  \citenamefont {Beach}}]{emori_current-driven_2013}%
  \BibitemOpen
  \bibfield  {author} {\bibinfo {author} {\bibfnamefont {Satoru}\ \bibnamefont
  {Emori}}, \bibinfo {author} {\bibfnamefont {Uwe}\ \bibnamefont {Bauer}},
  \bibinfo {author} {\bibfnamefont {Sung-Min}\ \bibnamefont {Ahn}}, \bibinfo
  {author} {\bibfnamefont {Eduardo}\ \bibnamefont {Martinez}}, \ and\ \bibinfo
  {author} {\bibfnamefont {Geoffrey S.~D.}\ \bibnamefont {Beach}},\ }\bibfield
  {title} {\enquote {\bibinfo {title} {Current-driven dynamics of chiral
  ferromagnetic domain walls},}\ }\href {\doibase 10.1038/nmat3675} {\bibfield
  {journal} {\bibinfo  {journal} {Nature Materials}\ }\textbf {\bibinfo
  {volume} {12}},\ \bibinfo {pages} {611--616} (\bibinfo {year}
  {2013})}\BibitemShut {NoStop}%
\bibitem [{\citenamefont {Ryu}\ \emph {et~al.}(2013)\citenamefont {Ryu},
  \citenamefont {Thomas}, \citenamefont {Yang},\ and\ \citenamefont
  {Parkin}}]{ryu_chiral_2013}%
  \BibitemOpen
  \bibfield  {author} {\bibinfo {author} {\bibfnamefont {Kwang-Su}\
  \bibnamefont {Ryu}}, \bibinfo {author} {\bibfnamefont {Luc}\ \bibnamefont
  {Thomas}}, \bibinfo {author} {\bibfnamefont {See-Hun}\ \bibnamefont {Yang}},
  \ and\ \bibinfo {author} {\bibfnamefont {Stuart}\ \bibnamefont {Parkin}},\
  }\bibfield  {title} {\enquote {\bibinfo {title} {Chiral spin torque at
  magnetic domain walls},}\ }\href {\doibase 10.1038/nnano.2013.102} {\bibfield
   {journal} {\bibinfo  {journal} {Nature Nanotechnology}\ }\textbf {\bibinfo
  {volume} {8}},\ \bibinfo {pages} {527--533} (\bibinfo {year}
  {2013})}\BibitemShut {NoStop}%
\bibitem [{\citenamefont {Emori}\ \emph {et~al.}(2014)\citenamefont {Emori},
  \citenamefont {Martinez}, \citenamefont {Lee}, \citenamefont {Lee},
  \citenamefont {Bauer}, \citenamefont {Ahn}, \citenamefont {Agrawal},
  \citenamefont {Bono},\ and\ \citenamefont {Beach}}]{PhysRevB.90.184427}%
  \BibitemOpen
  \bibfield  {author} {\bibinfo {author} {\bibfnamefont {Satoru}\ \bibnamefont
  {Emori}}, \bibinfo {author} {\bibfnamefont {Eduardo}\ \bibnamefont
  {Martinez}}, \bibinfo {author} {\bibfnamefont {Kyung-Jin}\ \bibnamefont
  {Lee}}, \bibinfo {author} {\bibfnamefont {Hyun-Woo}\ \bibnamefont {Lee}},
  \bibinfo {author} {\bibfnamefont {Uwe}\ \bibnamefont {Bauer}}, \bibinfo
  {author} {\bibfnamefont {Sung-Min}\ \bibnamefont {Ahn}}, \bibinfo {author}
  {\bibfnamefont {Parnika}\ \bibnamefont {Agrawal}}, \bibinfo {author}
  {\bibfnamefont {David~C.}\ \bibnamefont {Bono}}, \ and\ \bibinfo {author}
  {\bibfnamefont {Geoffrey S.~D.}\ \bibnamefont {Beach}},\ }\bibfield  {title}
  {\enquote {\bibinfo {title} {Spin hall torque magnetometry of dzyaloshinskii
  domain walls},}\ }\href {\doibase 10.1103/PhysRevB.90.184427} {\bibfield
  {journal} {\bibinfo  {journal} {Phys. Rev. B}\ }\textbf {\bibinfo {volume}
  {90}},\ \bibinfo {pages} {184427} (\bibinfo {year} {2014})}\BibitemShut
  {NoStop}%
\bibitem [{\citenamefont {Liu}\ \emph {et~al.}(2012)\citenamefont {Liu},
  \citenamefont {Pai}, \citenamefont {Li}, \citenamefont {Tseng}, \citenamefont
  {Ralph},\ and\ \citenamefont {Buhrman}}]{liu_spin-torque_2012}%
  \BibitemOpen
  \bibfield  {author} {\bibinfo {author} {\bibfnamefont {Luqiao}\ \bibnamefont
  {Liu}}, \bibinfo {author} {\bibfnamefont {Chi-Feng}\ \bibnamefont {Pai}},
  \bibinfo {author} {\bibfnamefont {Y.}~\bibnamefont {Li}}, \bibinfo {author}
  {\bibfnamefont {H.~W.}\ \bibnamefont {Tseng}}, \bibinfo {author}
  {\bibfnamefont {D.~C.}\ \bibnamefont {Ralph}}, \ and\ \bibinfo {author}
  {\bibfnamefont {R.~A.}\ \bibnamefont {Buhrman}},\ }\bibfield  {title}
  {\enquote {\bibinfo {title} {Spin-{Torque} {Switching} with the {Giant}
  {Spin} {Hall} {Effect} of {Tantalum}},}\ }\href {\doibase
  10.1126/science.1218197} {\bibfield  {journal} {\bibinfo  {journal}
  {Science}\ }\textbf {\bibinfo {volume} {336}},\ \bibinfo {pages} {555--558}
  (\bibinfo {year} {2012})}\BibitemShut {NoStop}%
\bibitem [{\citenamefont {Jiang}\ \emph {et~al.}(2015)\citenamefont {Jiang},
  \citenamefont {Upadhyaya}, \citenamefont {Zhang}, \citenamefont {Yu},
  \citenamefont {Jungfleisch}, \citenamefont {Fradin}, \citenamefont {Pearson},
  \citenamefont {Tserkovnyak}, \citenamefont {Wang}, \citenamefont {Heinonen},
  \citenamefont {te~Velthuis},\ and\ \citenamefont {Hoffmann}}]{Jiang17072015}%
  \BibitemOpen
  \bibfield  {author} {\bibinfo {author} {\bibfnamefont {Wanjun}\ \bibnamefont
  {Jiang}}, \bibinfo {author} {\bibfnamefont {Pramey}\ \bibnamefont
  {Upadhyaya}}, \bibinfo {author} {\bibfnamefont {Wei}\ \bibnamefont {Zhang}},
  \bibinfo {author} {\bibfnamefont {Guoqiang}\ \bibnamefont {Yu}}, \bibinfo
  {author} {\bibfnamefont {M.~Benjamin}\ \bibnamefont {Jungfleisch}}, \bibinfo
  {author} {\bibfnamefont {Frank~Y.}\ \bibnamefont {Fradin}}, \bibinfo {author}
  {\bibfnamefont {John~E.}\ \bibnamefont {Pearson}}, \bibinfo {author}
  {\bibfnamefont {Yaroslav}\ \bibnamefont {Tserkovnyak}}, \bibinfo {author}
  {\bibfnamefont {Kang~L.}\ \bibnamefont {Wang}}, \bibinfo {author}
  {\bibfnamefont {Olle}\ \bibnamefont {Heinonen}}, \bibinfo {author}
  {\bibfnamefont {Suzanne G.~E.}\ \bibnamefont {te~Velthuis}}, \ and\ \bibinfo
  {author} {\bibfnamefont {Axel}\ \bibnamefont {Hoffmann}},\ }\bibfield
  {title} {\enquote {\bibinfo {title} {Blowing magnetic skyrmion bubbles},}\
  }\href {\doibase 10.1126/science.aaa1442} {\bibfield  {journal} {\bibinfo
  {journal} {Science}\ }\textbf {\bibinfo {volume} {349}},\ \bibinfo {pages}
  {283--286} (\bibinfo {year} {2015})}\BibitemShut {NoStop}%
\bibitem [{\citenamefont {Jiang}\ \emph {et~al.}(2017)\citenamefont {Jiang},
  \citenamefont {Zhang}, \citenamefont {Yu}, \citenamefont {Zhang},
  \citenamefont {Wang}, \citenamefont {Benjamin Jungfleisch}, \citenamefont
  {Pearson}, \citenamefont {Cheng}, \citenamefont {Heinonen}, \citenamefont
  {Wang}, \citenamefont {Zhou}, \citenamefont {Hoffmann},\ and\ \citenamefont
  {te Velthuis}}]{jiang_direct_2017}%
  \BibitemOpen
  \bibfield  {author} {\bibinfo {author} {\bibfnamefont {Wanjun}\ \bibnamefont
  {Jiang}}, \bibinfo {author} {\bibfnamefont {Xichao}\ \bibnamefont {Zhang}},
  \bibinfo {author} {\bibfnamefont {Guoqiang}\ \bibnamefont {Yu}}, \bibinfo
  {author} {\bibfnamefont {Wei}\ \bibnamefont {Zhang}}, \bibinfo {author}
  {\bibfnamefont {Xiao}\ \bibnamefont {Wang}}, \bibinfo {author} {\bibfnamefont
  {M.}~\bibnamefont {Benjamin Jungfleisch}}, \bibinfo {author} {\bibfnamefont
  {John~E.}\ \bibnamefont {Pearson}}, \bibinfo {author} {\bibfnamefont
  {Xuemei}\ \bibnamefont {Cheng}}, \bibinfo {author} {\bibfnamefont {Olle}\
  \bibnamefont {Heinonen}}, \bibinfo {author} {\bibfnamefont {Kang~L.}\
  \bibnamefont {Wang}}, \bibinfo {author} {\bibfnamefont {Yan}\ \bibnamefont
  {Zhou}}, \bibinfo {author} {\bibfnamefont {Axel}\ \bibnamefont {Hoffmann}}, \
  and\ \bibinfo {author} {\bibfnamefont {Suzanne G.~E.}\ \bibnamefont
  {te Velthuis}},\ }\bibfield  {title} {\enquote {\bibinfo {title} {Direct
  observation of the skyrmion {Hall} effect},}\ }\href {\doibase
  10.1038/nphys3883} {\bibfield  {journal} {\bibinfo  {journal} {Nature
  Physics}\ }\textbf {\bibinfo {volume} {13}},\ \bibinfo {pages} {162--169}
  (\bibinfo {year} {2017})}\BibitemShut {NoStop}%
\bibitem [{\citenamefont {Heinze}\ \emph {et~al.}(2011)\citenamefont {Heinze},
  \citenamefont {Bergmann}, \citenamefont {Menzel}, \citenamefont {Brede},
  \citenamefont {Kubetzka}, \citenamefont {Wiesendanger}, \citenamefont
  {Bihlmayer},\ and\ \citenamefont {Bl��gel}}]{Heinze2011}%
  \BibitemOpen
  \bibfield  {author} {\bibinfo {author} {\bibfnamefont {Stefan}\ \bibnamefont
  {Heinze}}, \bibinfo {author} {\bibfnamefont {Kirsten~von}\ \bibnamefont
  {Bergmann}}, \bibinfo {author} {\bibfnamefont {Matthias}\ \bibnamefont
  {Menzel}}, \bibinfo {author} {\bibfnamefont {Jens}\ \bibnamefont {Brede}},
  \bibinfo {author} {\bibfnamefont {Andr��}\ \bibnamefont {Kubetzka}},
  \bibinfo {author} {\bibfnamefont {Roland}\ \bibnamefont {Wiesendanger}},
  \bibinfo {author} {\bibfnamefont {Gustav}\ \bibnamefont {Bihlmayer}}, \ and\
  \bibinfo {author} {\bibfnamefont {Stefan}\ \bibnamefont {Bl��gel}},\
  }\bibfield  {title} {\enquote {\bibinfo {title} {Spontaneous atomic-scale
  magnetic skyrmion lattice in two dimensions},}\ }\href {\doibase
  10.1038/nphys2045} {\bibfield  {journal} {\bibinfo  {journal} {Nature
  Physics}\ }\textbf {\bibinfo {volume} {7}},\ \bibinfo {pages} {713} (\bibinfo
  {year} {2011})}\BibitemShut {NoStop}%
\bibitem [{\citenamefont {Romming}\ \emph {et~al.}(2013)\citenamefont
  {Romming}, \citenamefont {Hanneken}, \citenamefont {Menzel}, \citenamefont
  {Bickel}, \citenamefont {Wolter}, \citenamefont {Bergmann}, \citenamefont
  {Kubetzka},\ and\ \citenamefont {Wiesendanger}}]{Romming2013}%
  \BibitemOpen
  \bibfield  {author} {\bibinfo {author} {\bibfnamefont {Niklas}\ \bibnamefont
  {Romming}}, \bibinfo {author} {\bibfnamefont {Christian}\ \bibnamefont
  {Hanneken}}, \bibinfo {author} {\bibfnamefont {Matthias}\ \bibnamefont
  {Menzel}}, \bibinfo {author} {\bibfnamefont {Jessica~E.}\ \bibnamefont
  {Bickel}}, \bibinfo {author} {\bibfnamefont {Boris}\ \bibnamefont {Wolter}},
  \bibinfo {author} {\bibfnamefont {Kirsten~von}\ \bibnamefont {Bergmann}},
  \bibinfo {author} {\bibfnamefont {Andr\'{e}}\ \bibnamefont {Kubetzka}}, \
  and\ \bibinfo {author} {\bibfnamefont {Roland}\ \bibnamefont
  {Wiesendanger}},\ }\bibfield  {title} {\enquote {\bibinfo {title} {Writing
  and deleting single magnetic skyrmions},}\ }\href {\doibase
  10.1126/science.1240573} {\bibfield  {journal} {\bibinfo  {journal}
  {Science}\ }\textbf {\bibinfo {volume} {341}},\ \bibinfo {pages} {636--639}
  (\bibinfo {year} {2013})}\BibitemShut {NoStop}%
\bibitem [{\citenamefont {Thiaville}\ \emph {et~al.}(2012)\citenamefont
  {Thiaville}, \citenamefont {Rohart}, \citenamefont {Ju\'{e}}, \citenamefont
  {Cros},\ and\ \citenamefont {Fert}}]{thiaville_dynamics_2012}%
  \BibitemOpen
  \bibfield  {author} {\bibinfo {author} {\bibfnamefont {Andr\'{e}}\
  \bibnamefont {Thiaville}}, \bibinfo {author} {\bibfnamefont {Stanislas}\
  \bibnamefont {Rohart}}, \bibinfo {author} {\bibfnamefont {\'{E}milie}\
  \bibnamefont {Ju\'{e}}}, \bibinfo {author} {\bibfnamefont {Vincent}\
  \bibnamefont {Cros}}, \ and\ \bibinfo {author} {\bibfnamefont {Albert}\
  \bibnamefont {Fert}},\ }\bibfield  {title} {\enquote {\bibinfo {title}
  {Dynamics of {Dzyaloshinskii} domain walls in ultrathin magnetic films},}\
  }\href {\doibase 10.1209/0295-5075/100/57002} {\bibfield  {journal} {\bibinfo
   {journal} {EPL (Europhysics Letters)}\ }\textbf {\bibinfo {volume} {100}},\
  \bibinfo {pages} {57002} (\bibinfo {year} {2012})}\BibitemShut {NoStop}%
\bibitem [{\citenamefont {Makhfudz}\ \emph {et~al.}(2012)\citenamefont
  {Makhfudz}, \citenamefont {Kr\"uger},\ and\ \citenamefont
  {Tchernyshyov}}]{Makhfudz2012}%
  \BibitemOpen
  \bibfield  {author} {\bibinfo {author} {\bibfnamefont {Imam}\ \bibnamefont
  {Makhfudz}}, \bibinfo {author} {\bibfnamefont {Benjamin}\ \bibnamefont
  {Kr\"uger}}, \ and\ \bibinfo {author} {\bibfnamefont {Oleg}\ \bibnamefont
  {Tchernyshyov}},\ }\bibfield  {title} {\enquote {\bibinfo {title} {Inertia
  and chiral edge modes of a skyrmion magnetic bubble},}\ }\href {\doibase
  10.1103/PhysRevLett.109.217201} {\bibfield  {journal} {\bibinfo  {journal}
  {Phys. Rev. Lett.}\ }\textbf {\bibinfo {volume} {109}},\ \bibinfo {pages}
  {217201} (\bibinfo {year} {2012})}\BibitemShut {NoStop}%
\bibitem [{\citenamefont {Okubo}\ \emph {et~al.}(2012)\citenamefont {Okubo},
  \citenamefont {Chung},\ and\ \citenamefont
  {Kawamura}}]{PhysRevLett.108.017206}%
  \BibitemOpen
  \bibfield  {author} {\bibinfo {author} {\bibfnamefont {Tsuyoshi}\
  \bibnamefont {Okubo}}, \bibinfo {author} {\bibfnamefont {Sungki}\
  \bibnamefont {Chung}}, \ and\ \bibinfo {author} {\bibfnamefont {Hikaru}\
  \bibnamefont {Kawamura}},\ }\bibfield  {title} {\enquote {\bibinfo {title}
  {Multiple-$q$ states and the skyrmion lattice of the triangular-lattice
  heisenberg antiferromagnet under magnetic fields},}\ }\href {\doibase
  10.1103/PhysRevLett.108.017206} {\bibfield  {journal} {\bibinfo  {journal}
  {Phys. Rev. Lett.}\ }\textbf {\bibinfo {volume} {108}},\ \bibinfo {pages}
  {017206} (\bibinfo {year} {2012})}\BibitemShut {NoStop}%
\bibitem [{\citenamefont {Leonov}\ and\ \citenamefont
  {Mostovoy}(2015)}]{leonov_multiply_2015}%
  \BibitemOpen
  \bibfield  {author} {\bibinfo {author} {\bibfnamefont {A.~O.}\ \bibnamefont
  {Leonov}}\ and\ \bibinfo {author} {\bibfnamefont {M.}~\bibnamefont
  {Mostovoy}},\ }\bibfield  {title} {\enquote {\bibinfo {title} {Multiply
  periodic states and isolated skyrmions in an anisotropic frustrated
  magnet},}\ }\href {\doibase 10.1038/ncomms9275} {\bibfield  {journal}
  {\bibinfo  {journal} {Nature Communications}\ }\textbf {\bibinfo {volume}
  {6}},\ \bibinfo {pages} {8275} (\bibinfo {year} {2015})}\BibitemShut
  {NoStop}%
\bibitem [{\citenamefont {Lin}\ and\ \citenamefont
  {Hayami}(2016)}]{PhysRevB.93.064430}%
  \BibitemOpen
  \bibfield  {author} {\bibinfo {author} {\bibfnamefont {Shi-Zeng}\
  \bibnamefont {Lin}}\ and\ \bibinfo {author} {\bibfnamefont {Satoru}\
  \bibnamefont {Hayami}},\ }\bibfield  {title} {\enquote {\bibinfo {title}
  {Ginzburg-landau theory for skyrmions in inversion-symmetric magnets with
  competing interactions},}\ }\href {\doibase 10.1103/PhysRevB.93.064430}
  {\bibfield  {journal} {\bibinfo  {journal} {Phys. Rev. B}\ }\textbf {\bibinfo
  {volume} {93}},\ \bibinfo {pages} {064430} (\bibinfo {year}
  {2016})}\BibitemShut {NoStop}%
\bibitem [{\citenamefont {Hayami}\ \emph {et~al.}(2016)\citenamefont {Hayami},
  \citenamefont {Lin},\ and\ \citenamefont {Batista}}]{PhysRevB.93.184413}%
  \BibitemOpen
  \bibfield  {author} {\bibinfo {author} {\bibfnamefont {Satoru}\ \bibnamefont
  {Hayami}}, \bibinfo {author} {\bibfnamefont {Shi-Zeng}\ \bibnamefont {Lin}},
  \ and\ \bibinfo {author} {\bibfnamefont {Cristian~D.}\ \bibnamefont
  {Batista}},\ }\bibfield  {title} {\enquote {\bibinfo {title} {Bubble and
  skyrmion crystals in frustrated magnets with easy-axis anisotropy},}\ }\href
  {\doibase 10.1103/PhysRevB.93.184413} {\bibfield  {journal} {\bibinfo
  {journal} {Phys. Rev. B}\ }\textbf {\bibinfo {volume} {93}},\ \bibinfo
  {pages} {184413} (\bibinfo {year} {2016})}\BibitemShut {NoStop}%
\bibitem [{\citenamefont {Zhang}\ \emph {et~al.}(2017)\citenamefont {Zhang},
  \citenamefont {Xia}, \citenamefont {Zhou}, \citenamefont {Liu},\ and\
  \citenamefont {Ezawa}}]{zhang_skyrmions_2017}%
  \BibitemOpen
  \bibfield  {author} {\bibinfo {author} {\bibfnamefont {Xichao}\ \bibnamefont
  {Zhang}}, \bibinfo {author} {\bibfnamefont {Jing}\ \bibnamefont {Xia}},
  \bibinfo {author} {\bibfnamefont {Yan}\ \bibnamefont {Zhou}}, \bibinfo
  {author} {\bibfnamefont {Xiaoxi}\ \bibnamefont {Liu}}, \ and\ \bibinfo
  {author} {\bibfnamefont {Motohiko}\ \bibnamefont {Ezawa}},\ }\bibfield
  {title} {\enquote {\bibinfo {title} {Skyrmions and {Antiskyrmions} in a
  {Frustrated} ${J}_1$-${J}_2$-${J}_3$ {Ferromagnetic} {Film}:
  {Current}-induced {Helicity} {Locking}-{Unlocking} {Transition}},}\ }\href
  {http://arxiv.org/abs/1703.07501} {\bibfield  {journal} {\bibinfo  {journal}
  {arXiv:1703.07501 [cond-mat]}\ } (\bibinfo {year} {2017})},\ \bibinfo {note}
  {arXiv: 1703.07501}\BibitemShut {NoStop}%
\bibitem [{\citenamefont {R\'ozsa}\ \emph {et~al.}(2017)\citenamefont
  {R\'ozsa}, \citenamefont {Palot\'as}, \citenamefont {De\'ak}, \citenamefont
  {Simon}, \citenamefont {Yanes}, \citenamefont {Udvardi}, \citenamefont
  {Szunyogh},\ and\ \citenamefont {Nowak}}]{PhysRevB.95.094423}%
  \BibitemOpen
  \bibfield  {author} {\bibinfo {author} {\bibfnamefont {Levente}\ \bibnamefont
  {R\'ozsa}}, \bibinfo {author} {\bibfnamefont {Kriszti\'an}\ \bibnamefont
  {Palot\'as}}, \bibinfo {author} {\bibfnamefont {Andr\'as}\ \bibnamefont
  {De\'ak}}, \bibinfo {author} {\bibfnamefont {Eszter}\ \bibnamefont {Simon}},
  \bibinfo {author} {\bibfnamefont {Rocio}\ \bibnamefont {Yanes}}, \bibinfo
  {author} {\bibfnamefont {L\'aszl\'o}\ \bibnamefont {Udvardi}}, \bibinfo
  {author} {\bibfnamefont {L\'aszl\'o}\ \bibnamefont {Szunyogh}}, \ and\
  \bibinfo {author} {\bibfnamefont {Ulrich}\ \bibnamefont {Nowak}},\ }\bibfield
   {title} {\enquote {\bibinfo {title} {Formation and stability of metastable
  skyrmionic spin structures with various topologies in an ultrathin film},}\
  }\href {\doibase 10.1103/PhysRevB.95.094423} {\bibfield  {journal} {\bibinfo
  {journal} {Phys. Rev. B}\ }\textbf {\bibinfo {volume} {95}},\ \bibinfo
  {pages} {094423} (\bibinfo {year} {2017})}\BibitemShut {NoStop}%
\end{thebibliography}%

\end{document}